# Impermeable Inorganic "Walls" Sandwiching Photoactive Layer toward Inverted Perovskite Solar and Indoor−Photovoltaic Devices

*Jie Xu,[1,#] Jun Xi,[2,#,*] Hua Dong,[1,7,*] Namyoung Ahn,[4] Zonglong Zhu,[5] Jinbo Chen,[1] Peizhou Li,[1] Xinyi zhu,[1] Jinfei Dai,[1] Ziyang Hu,[6] Bo Jiao,[1] Xun Hou,[1] Jingrui Li,[3,*] and Zhaoxin Wu[1,7,*]*

[1]Key Laboratory for Physical Electronics and Devices of the Ministry of Education & Shaanxi Key Lab of Information Photonic Technique, School of Electronic Science and Engineering, Xi'an Jiaotong University, No.28, Xianning West Road, Xi'an, 710049, China.

[2]Zernike Institute for Advanced Materials, University of Groningen, Nijenborgh 4, 9747 AG Groningen, the Netherlands

[3]Electronic Materials Research Laboratory, Key Laboratory of the Ministry of Education & International Center for Dielectric Research, School of Electronic Science and Engineering, Xi'an Jiaotong University, Xi'an 710049, China.

[4]Chemistry Division, Los Alamos National Laboratory, Los Alamos, NM 87545, USA.

[5]Department of Chemistry, City University of Hong Kong, Kowloon, Hong Kong.

[6]Department of Microelectronic Science and Engineering, Ningbo University, Ningbo 315211, China.

[7]Collaborative Innovation Center of Extreme Optics, Shanxi University, Taiyuan 030006, China.

**[#]These two authors contributed equally to this work.**
***Corresponding authors.**
**E-mail:**
**j.xi@rug.nl**
**donghuaxjtu@mail.xjtu.edu.cn**
**jingrui.li@mail.xjtu.edu.cn**
**zhaoxinwu@mail.xjtu.edu.cn**

## Abstract

Interfaces between the perovskite active layer and the charge-transport layers (CTLs) play a critical role in both efficiency and stability of halide-perovskite photovoltaics. One of the major concerns is that surface defects of perovskite could cause detrimental nonradiative recombination and material degradation. In this work, we addressed this challenging problem by inserting ultrathin alkali-fluoride (AF) films between the tri-cation lead-iodide perovskite layer and both CTLs. This bilateral inorganic "walls" strategy makes use of both physical-blocking and chemical-anchoring functionalities of the continuous, uniform and compact AF framework: on the one hand, the uniformly distributed alkali-iodine coordination at the perovskite-AF interfaces effectively suppresses the formation of iodine-vacancy defects at the surfaces and grain boundaries of the whole perovskite film, thus reducing the trap-assisted recombination at the perovskite-CTL interfaces and therewith the open-voltage loss; on the other hand, the impermeable AF buffer layers effectively prevent the bidirectional ion migration at the perovskite-CTLs interfaces even under harsh working conditions. As a result, a power-conversion efficiency (PCE) of 22.02% (certified efficiency 20.4%) with low open-voltage deficit (< 0.4V) was achieved for the low-temperature processed inverted planar perovskite solar cells. Exceptional operational stability (500 h, ISOS-L-2) and thermal stability (1000 h, ISOS-D-2) were obtained. Meanwhile, a 35.7% PCE was obtained under dim-light source (1000 lux white LED light) with the optimized device, which is among the best records in perovskite indoor photovoltaics.

## Introduction

In recent years, people have witnessed an astonishing progress of organic-inorganic hybrid perovskite optoelectronic devices thanks to their excellent electronic, optical, and charge-transport properties.[1-7] Currently, the power conversion efficiency (PCE) of perovskite solar cells (PSCs) and perovskite indoor-photovoltaic devices have reached 25.5% and 36%, respectively, rivaling the traditional silicon- and GaAs-based devices.[8-9] Nevertheless, there is still a long way to go for approaching the Shockley-Queisser theoretical limit (34% under AM 1.5G solar spectrum) of single-junction devices.[10] So far, a majority of achievements that simultaneously realized high efficiency and high stability are based on the n-i-p (normal) device structure, for which the thick hole-transport layer (HTL) with chemical doping leads to high fabrication cost and inevitable stability risk. In contrast, PSCs of inverted (p-i-n) planar structure with low processing temperature and thinner, undoped HTL are considered promising for industrial production of versatile devices ranging from standard single-junction cells to flexible and tandem cells.[11-15] The latest record of inverted-structure PSC's PCE is 23.0%.[12]

Unfortunately, inverted-structure PSCs still lag behind the n-i-p counterparts in terms of both efficiency and stability. Nonradiative recombination induced by defects in the bulk phase of material and at the surfaces and grain boundaries is one of the main factors that limits the device performance.[11,16] While bulk defects can be properly controlled by means of several mature methodologies,[17-21] suppressing the formation of defects at the perovskite surfaces remains a major challenge in today's perovskite optoelectronic community.[11,16] Specifically, these defects especially iodine-vacancy ($v_I$) induce trap states within the bandgap of the perovskite semiconductor, which facilitate nonradiative recombination at the interfaces thus giving rise to open-circuit voltage ($V_{OC}$) loss and therewith limited PCE.[12] In addition, the device stability could be strongly affected by the trapped charges at the interfaces between the perovskite layer and CTLs upon the defect formation, as the thus-induced electric field would initiate and accelerate the decomposition of perovskite and deprotonation of the organic moieties.[22-23]

A variety of methods have been developed to mediate the interfacial problem, which can be classified into the organic and the inorganic strategies. Amine-group-containing large organic ligands with a favorable diversity of candidate molecules have proven to be effective in passivating the perovskite surfaces.[12,24-25] However, the regular orientation of long-chain or bulky organic molecules is hard to achieved on a length scale beyond a few lattice constants of perovskite. Hence, steric hindrance within the organic-ligand layer is prevailing and problematic, as it induces the distortion of perovskite lattices at the surface as well as the decay and incomplete coverage of the organic layer, which eventually leads to the instability of materials and devices in long-term applications under harsh working conditions. The inorganic strategy includes two branches: alkali-cation doping and inorganic buffer-layer introduction. modification system is also reported benefiting from its own stability. It was widely reported that incorporating $Li^+$, $Na^+$, $K^+$, and $Rb^+$ into the perovskite lattice can significantly reduce the nonradiative recombination thus improving the efficiency and stability.[26-30] However, how to guarantee that the doping takes effect uniformly and completely at the perovskite surfaces and grain boundaries remains a problem. In addition, further surface passivation (such as via introducing organic ligands) might be required, as the surfaces of the doped perovskite is not yet physically well-protect against interfacial ion diffusion and possible contact with environmental small molecules such as oxygen and water. In parallel, inorganic buffer-layer technologies are also employed to inhibit the improper ion diffusion.[31,32] It is required that the interaction between the buffer layer and the perovskite can anchor the surface atoms of the latter without inducing a large perovskite-lattice distortion. The choice of inorganic materials is important, however, corresponding discussion about the choice of inorganic materials is still rare. Thus an inorganic protocol designed to simultaneously minimize the defects and block the ion diffusion over the interfaces between the perovskite layer and both CTLs in the inverted planar devices is highly appealed.

In this work, we developed an inorganic bilateral strategy to engineer the surfaces of the tri-cation lead-iodide perovskite layer in the inverted planar structure of PSC. Ultrathin alkali-halide buffer layers were facilely inserted between the perovskite layer and both the electron-transport layer (ETL) and the HTL. Considering F's smallest anionic radius and therewith the most stable and compact alkali-fluoride (AF) frameworks, we used fluorides with different alkali cations as buffer-layer materials in this work, namely lithium fluoride (LiF), sodium fluoride (NaF), and potassium fluoride (KF). Benefiting from the chemical-anchoring functionality, the uniformly distributed alkali-iodide coordination between the buffer layer and perovskite could effectively passivate the surface defects without affecting the perovskite lattice. Moreover, the bilateral protection could physically prevent the bidirectional ion migration and oxygen or moisture invasion to the perovskite due to AF's compact and stable features. With this facile bilateral inorganic "walls" strategy, we achieved a PCE of 22.02% for a KF-modified inverted planar solar cell, which allows an exceptional operating stability (500 h, ISOS-L-2) and thermal stability (1000 h, ISOS-D-2). Interestingly, the optimized bilaterally-modified device also exhibited a remarkable performance under dim-light condition, which achieved a 35.7% efficiency under 1000 lux white LED light. We expect this work can provide an inorganic perspective to effectively manage the trap-assisted defects and facilitate long-term stable and highly efficient perovskite optoelectronic devices toward industrial production.

# Results and discussion

## Device performance

The inverted planar perovskite-based device with bilateral AF modification (denoted by AF-perovskite-AF hereafter) is sketched in **Figure 1A**. The traditional one-step antisolvent method was adopted to prepare the perovskite thin film with optimized component $Cs_{0.05}(FA_{0.88}MA_{0.12})_{0.95}PbI_3$ on the substrate (details given in Supplemental Information). The cross-sectional scanning electron microscopy (SEM) image in **Figure 1B** shows a perfect planar device structure, which comprises stacked layers of indium tin oxide (ITO, 130 nm) / poly[bis(4-phenyl)(2,4,6-trimethylphenyl)amine] (PTAA, 8 nm) / perovskite (500 nm) / $C_{60}$ (30 nm) / bathocuproine (BCP, 6 nm) / Ag (120 nm). Notably, the bilateral AF buffer layers are too thin (< 5 nm) to observe.

The dependence of device performance on respective AF thickness is summarized in **Table S1**. For A = Li, Na, or K, the device efficiency was maximized with the thickness of both buffer layers equal to 3 nm. The statistical distributions of PCE and $V_{OC}$ from 120 devices with or without modification are shown in **Figures 1C** and **D**, respectively. The representative PCE for the control (i.e., without bilateral AF modification), LiF-perovskite-LiF, NaF-perovskite-NaF, and KF-perovskite-KF devices are 18.45%, 19.76%, 20.22%, and 21.29%, respectively. This finding indicates that the bilateral AF "walls" significantly improve the photovoltaic performance of devices, and exhibits the trend of increasing PCE from LiF, NaF to KF. This is mainly contributed by the significant enhancement of $V_{OC}$ (from 1.06 V for the control to 1.15 V for the KF-perovskite-KF device), while other metrics were also slightly improved (**Figure S1**). **Figure 1E** gives the current density – voltage (*J-V*) curves of the best-performing control and KF-perovskite-KF devices. The performance parameters of control are PCE = 19.20%, short circuit current density ($J_{SC}$) = 22.56 mA/cm$^2$, $V_{OC}$ = 1.067 V, and fill factor (FF) = 79.80%; while the KF-perovskite-KF device achieves a largely improved PCE of 22.02% with $J_{SC}$ = 23.40 mA/cm$^2$, $V_{OC}$ = 1.152 V, and FF = 81.70%. The parameters of other two AF-modified devices are given in **Table S2** and **Figure S2**. To the best of our knowledge, we have achieved the highest efficiency of inverted planar devices with a simple buffer-layer technology (detailed reference data given in **Table S3**). To confirm our in-laboratory device efficiency measurements, external certification of our nonencapsulated devices by the National Institute of Metrology, China, shows a PCE of 20.4% ($V_{OC}$ = 1.094 V, $J_{SC}$ = 23.22 mA/cm$^2$, and FF = 80.2%) as given in **Figure S3.** Meanwhile, the steady-state photocurrents of the best-performing devices at maximum power point tracking (inset of **Figure 1E**) are in good agreement with the *J-V* scanning results, thus supporting the reliability of the highest efficiency. As shown in **Figure S4**, the integrated short-circuit current densities from the incident photon-to-electron conversion efficiencies (IPCE) are consistent with the $J_{SC}$ from *J-V* measurement, thus validating the measured device performance. **Figure S5A** shows the representative *J-V* curves of the KF-perovskite-KF device and the control device under the reverse and forward scan, whose performance parameters are listed in **Table S4**. Using hysteresis index (HI) which is defined by HI = [PCE(reverse) - PCE(forward)] / PCE (forward) to characterize the hysteresis, **Table S4** indicates that the bilateral strategy with KF has largely reduced the HI from 3.3% to 0.3%, thus signifying perfectly balanced charge carrier extraction. For the KF-perovskite-KF device with 1 $cm^2$ solar cells, **Figure S5B** shows a PCE of 19.65% with negligible hysteresis.

To characterize the stability improvement upon the bilateral KF-modification, we conducted a series of measurements on unencapsulated devices. The results of stability tested at 85 °C in a nitrogen atmosphere (ISOS-D-2) and under continuous one-sun illumination (ISOS-L-2) are shown in **Figures 1F** and **G**, respectively. From **Figure 1F**, the introduction of bilateral KF ultrathin buffer layers has significantly improved the thermal stability of devices, as they retain 90% of the initial PCE after 1000 h, while the control ones retain only 80%. **Figure 1G** indicates the outstanding light-stability of the KF-perovskite-KF devices as they retain 100% of the original PCE after being exposed to one-sun illumination for 500 h.

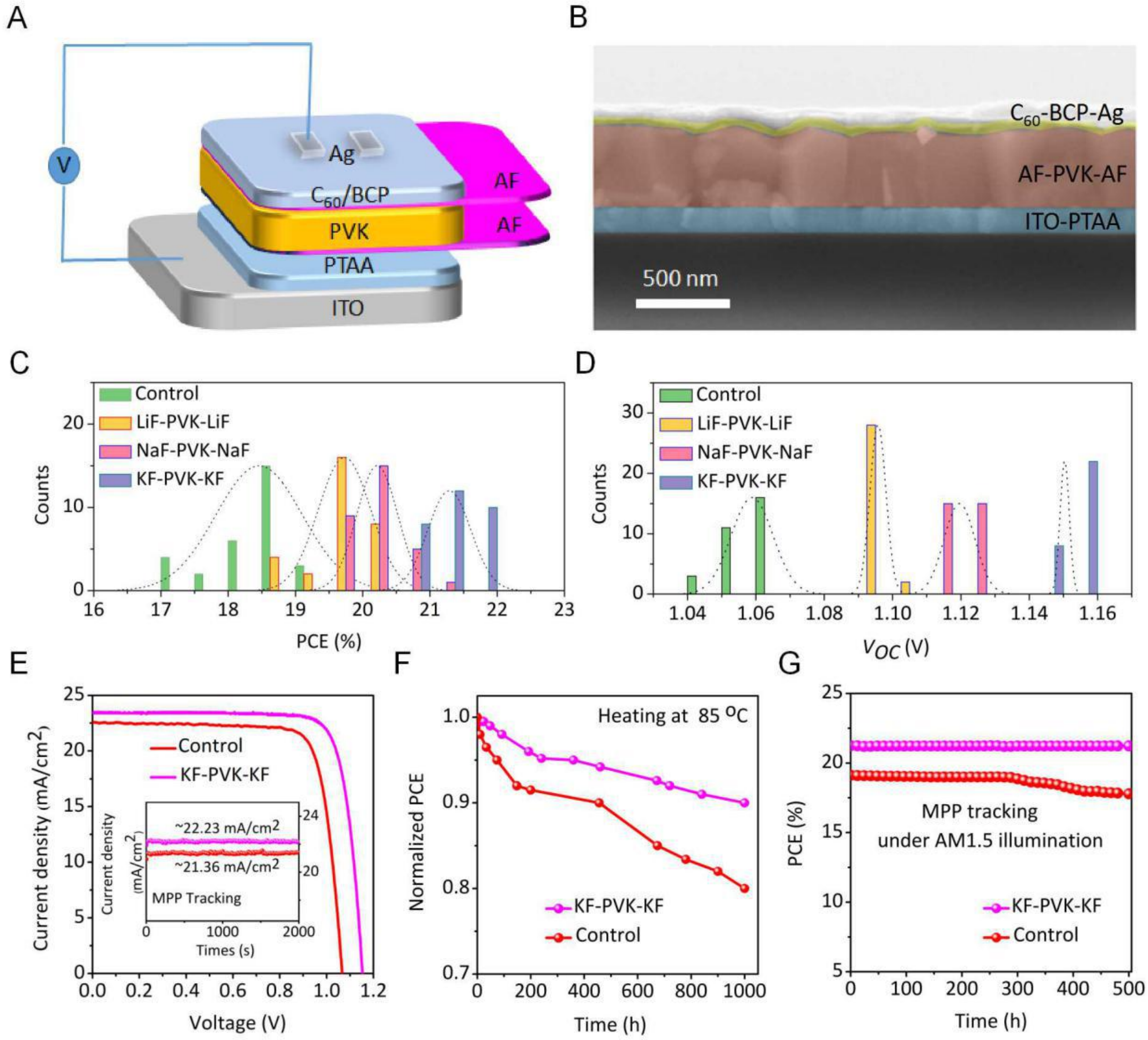

**Figure 1** (A) Schematic illustration of the PSCs device structure. (B) Cross-sectional SEM image of the actual device with different layers indicated by false-coloring (scale bar: 500 nm). (C) Statistical distribution of PCE for 120 devices (30 per structure). (D) Statistical distribution of $V_{OC}$ for 120 devices (30 per structure). (E) $J$-$V$ curves of champion devices (inset: stabilized photocurrent output at the maximum power point). Stabilities of the control and the bilateral KF-modified PSCs: (F) thermal stability by putting the PSCs at 85 °C hot plate in $N_2$ for 1000 h, and (G) operational stability of PSCs under continuous illumination in $N_2$ for 500 h. PVK stands for perovskite in the figures of this manuscript.

## Materials characterization of perovskite films and trap-state analysis

To understand the underlying physics of the performance improvement of the inverted planar PSCs upon the incorporation of bilateral AF (in particular KF) thin films into the device, we first characterize the related materials properties in this section, then analyze the mechanism of charge properties and processes in the next sections.

**Figure 2A** shows the morphology of the control and the KF-modified perovskite films. The blurred image in **Figure 2A**-right vividly shows that the continuous, ultrathin and transparent KF buffer layer almost completely covers the perovskite. One can observe that the KF modification results in a slightly larger grain size, which is possibly because the interaction with KF subtly mediates the surface tension of perovskite so as to affect the dynamics of nucleation.[33] Similar morphological features can be observed from the other films modified by the bilateral LiF and NaF (**Figure S6**). The profile elemental (alkali elements and fluorine) mapping (**Figure S7**) from the energy-dispersive X-ray (EDX) spectroscopy clearly indicates the homogeneous distribution of these elements on the perovskite surface, thus indicating a complete coverage of the AF buffer layers on the perovskite. **Figure S8** indicates that the element-content ratio of Pb to I is > 3 in all samples, which is consistent with the 10 wt% excess of $PbI_2$ in the precursor solution. Meanwhile, X-ray diffraction (XRD) measurement (**Figure S9**) shows higher crystallinity in all AF-modified perovskite films than the control one. Interestingly, a close view to the (100) peak (in terms of the quasi-cubic lattice of perovskite) of XRD (**Figure 2B**) illustrates a shift of this peak toward lower diffraction angle by about 0.1° upon the KF modification which corresponds to a ~ 0.7% expansion of the lattice constants. Previous studies suggested that such a lattice expansion may result from the occupancy of alkali cations at the interstitial sites,[29] which is evidenced by our X-ray photoelectron spectra (XPS) measurements (**Figure S10**). These findings indicate that a large number of alkali cations may enter the perovskite lattice at the surface region so that they can reinforce the surface iodide anions via ionic bonding. Besides, ultraviolet-visible absorption and photoluminescence (PL) spectra (**Figure S11A**) shows that the bilateral AF modification has a negligible effect on the bandgap (1.55 eV) of the perovskite samples. Overall, the changes of structural and optical properties of the perovskite bulk upon the incorporation of bilateral AF "walls" are relatively small.

The primary factors that give rise to large $V_{OC}$ loss thus restricting the PCE of PSCs include (a) nonradiative recombination in the bulk of the perovskite absorber, (b) unfavorable energy-level alignment at the perovskite-CTL interfaces, and (c) defect-induced (Shockley-Read-Hall or SRH) recombination at the perovskite surfaces and grain boundaries.[34] The only slight difference in the structural and optical properties between the AF-modified and the control samples, as already alluded to, implies that the obvious improvement of the performance parameters of PSCs upon the KF modification is mainly due to the suppressed SRH recombination at the perovskite surfaces namely its interfaces toward both CTLs. To confirm it, we first carried out PL-mapping and time-resolved PL lifetime (TRPL) mapping measurements that can intuitively describe the spatial distribution of radiative recombination events in both the control and the AF-modified perovskite films on the nanoscale. Consistent with the PL measurements (**Figure S11B**), PL-mapping (**Figure 2C**) indicates an overall stronger PL of KF-modified perovskite sample than the control. **Figure 2D** illustrates the distribution of normalized PL intensities, showing an obviously higher PL homogeneity in the KF-modified perovskite film. The lifetime distribution of TRPL-mapping (**Figure 2E**) reveals that the KF-perovskite-KF film is associated with a much longer carrier lifetime (395 ns) than the control (89 ns). TRPL-mapping and its statistics for the other two AF-modified films are given in **Figure S12**. The carrier lifetime follows the overall trend: control < LiF-perovskite-LiF < NaF-perovskite-NaF < KF-perovskite-KF, which is consistent with the trend of PSC performance. To summarize, both more pronounced radiative recombination as evidenced by the PL mapping, and the significantly longer charge-carrier lifetime given by the TRPL mapping, support that the introduction of double KF "walls" effectively restricts the nonradiative recombination at the perovskite surfaces and grain boundaries.

Thermal admittance spectroscopy (TAS) was carried out to characterize the local defect density of states (t-DOS). The experiment details and calculations of the t-DOS are given in Supplemental Information.[35,36] As shown in **Figure 2F** (and **Figure S13)**, a density of defect states on the order of $10^{18}$ $m^{-3}$ was estimated in the control sample. With the bilateral KF modification, the t-DOS is lower by nearly one order of magnitude. We can thus conclude that the introduction of KF ultrathin interlayers can efficiently suppress the formation of surface defects therewith inhibiting interfacial nonradiative recombination.

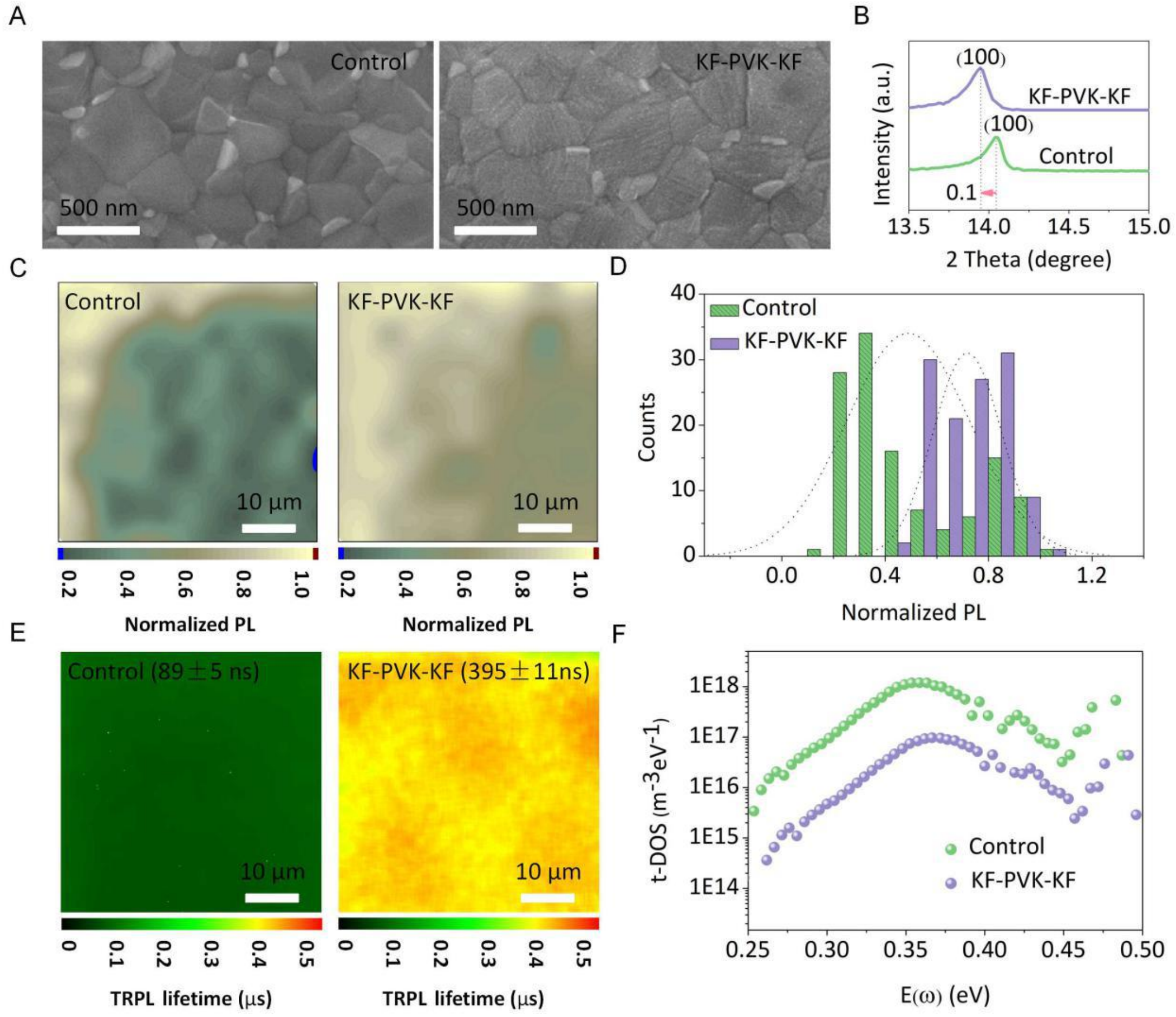

**Figure 2** (A) SEM images of the control and the bilateral KF-interlayers-modified perovskite films. (B) Close view of XRD spectra around the (100) peak. (C) PL mapping of the control and the KF-PVK-KF perovskite film (scanning: 50 μm × 50 μm ). (D) Statistic distribution of normalized PL for control and the KF-PVK-KF perovskite film. (E) TRPL-Mapping for control and the KF-PVK-KF perovskite film (scanning: 50 μm × 50 μm). (F) Trap density of states (t-DOS) for control and AF modified devices.

## Charge behaviors at interface

To further confirm that nonradiative recombination was largely reduced by introducing the bilateral AF "walls" into the system, we carried out a series of measurements on both ITO/HTL/perovskite (HP, as control) and ITO/HTL/interlayer/perovskite/interlayer (HIPI) structure (**Figure 3A**), with KF as the interlayer material due to the best performance of the corresponding PSC. **Figure S14** shows the topography of HP and HIPI measured by atomic force microscopy (AFM). Both samples are characterized by a root-mean-square roughness of 15 nm, which agrees with the observation from SEM images that the KF modification has not induced drastic changes in the morphology of the perovskite film. **Figure 3C** shows the contact potential difference (CPD) distribution of HP and HIPI, respectively, which was measured by Kelvin probe force microscopy (KPFM) under illumination. It is known that the value of CPD is related to the energy barrier of charge extraction at the interface.[37,38] A much lower CPD distribution is observed in HIPI than in HP (**Figure 3D**), reflecting a reduced charge-extraction energy barrier. Here the improved charge-extraction efficiency is attributed to the suppressed nonradiative recombination benefitting from the defect-passivating AF buffer layers. The analysis of CPD agrees well with the PL-mapping and TRPL-mapping. The decreased charge-extraction energy barrier is also closely related to the smaller $V_{OC}$ deficit.

To explore the functions of the inorganic AF interlayers, we first measured the PL with a simplified device structure that consists of two Au electrodes and the control or the KF-perovskite-KF film deposited on glass. With a DC voltage of 10 V, balanced charges of a high concentration were injected into the perovskite film where they recombined and emitted photons. The luminescence intensities of both samples were recorded as a function of injection time. **Figure 3E** shows that, at a short injection time (10 s), the luminescence intensities of the control and the KF-sandwiched perovskite films are relatively low and very similar. For a longer injection time (370 s), the control device emitted very much amplified light especially at the wavelengths near the optical band gap, while the intensity of KF-perovskite-KF increased only slightly. This is because the defect-induced states that can trap the injected charge carriers are much fewer in KF-perovskite-KF than in the control sample.

Overall, the charge-transport mechanism within the whole PSC device is illustrated by **Figure 3B**: the inserted ultrathin AF buffer layers effectively deactivate the unfavorable hole injection from perovskite to ETL ($C_{60}$/BCP) and the undesired electron injection from perovskite to HTL (PTAA), thus successfully minimizing the nonradiative recombination occurring at both perovskite-ETL and perovskite-HTL interfaces. The desired charge transfer at both interfaces is well facilitated via a reduced energy barrier, so that a reduced $V_{OC}$ loss and thus a high PCE of the photovoltaic device is guaranteed.

This approach is also expected to be effective to break the efficiency bottleneck in other optoelectronic devices, like light-emitting diodes (LEDs). To prove this potential, we fabricated quantum-dot (QD) LED device with the configuration ITO / AF / perovskite QDs / 2,2',2 -(1,3,5-Benzinetriyl)-tris(1-phenyl-1-H-benzimidazole) (TPBI) / AF / Al, as shown in **Figure S15B**, with the ITO/QD/TPBI/Al device (**Figure S15A**) as a reference, the current density and luminance of which are shown in **Figures S15 C** and **D**. Encouragingly, enhanced current density and luminance are found by the novel structure compared with the reference device, indicating improved carrier injection efficiency assisted by the bilateral AF buffer layers.

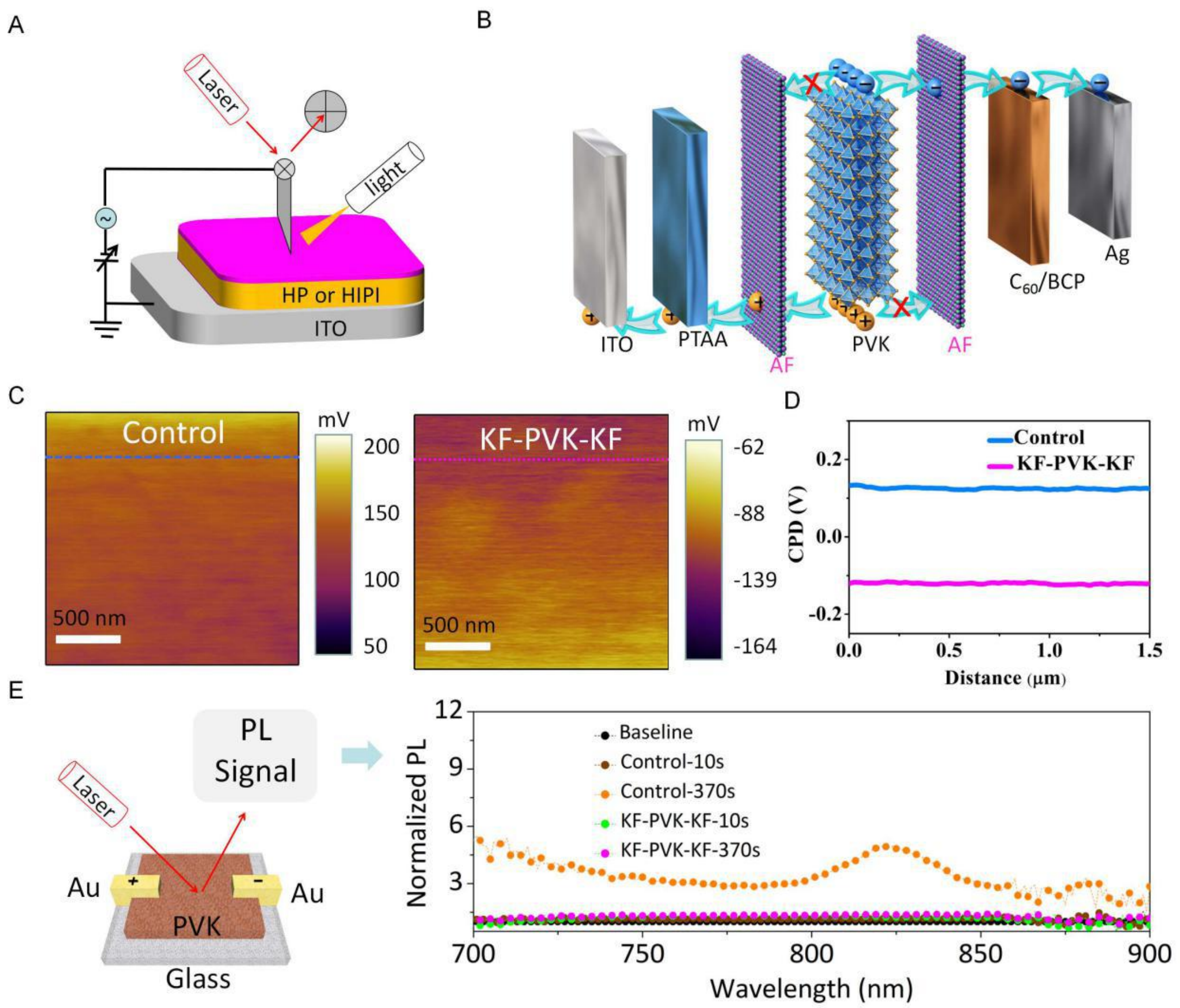

**Figure 3 (A)** Illustration of the KPFM setup. (B) Schematic diagram of charge-carriers extraction in the AF-PVK-AF based device. (C) Surface potential images of PVK/PTAA/ITO (control) and KF-PVK-KF-PTAA-ITO (KF-PVK-KF) sample from KPFM measurements under illumination conditions. (D) 1D line profiles of control and KF-PVK-KF under illumination. (E) Change of normalized PL intensity for control and KF-PVK-KF films under DC 10 V voltage sustaining for 10 s and 370 s.

## Mechanism of the chemical anchoring and physical blocking

For a theoretical insight of how an ultrathin AF protective coating passivates the surface defects of perovskite, we carried out density-functional-theory (DFT) calculations (details given in Supplemental Information) for a series of systems using the all-electron numeric-atom-centered orbital code FHI-aims.[39,40] Considering the relatively large model systems and the computational capacity, we constructed four-layer AF films to mimic the ultrathin interlayers (atomic structures shown in **Figure 4A**). Even so, the model systems for the $v_I$ calculations of the LiF-, NaF-, and KF-coated systems contain 1455, 1279, and 1175 atoms, respectively, all above the normal 1000-atom upper limit of traditional DFT calculations.

Two observables are of particular interest: the formation energy of each perovskite-AF complex and the formation energy of $v_I$ defect in each system. The DFT-calculated formation energies of perovskite-LiF, perovskite-NaF, and perovskite-KF are -0.270 eV, -0.109 eV, and -0.313 eV per surface Pb atom, respectively. All values are negative, indicating that the AF films can be spontaneously deposited to the perovskite surface to form stable complexes, thus protecting the perovskite material against possible contact with external particles (such as oxygen, water, and ions from the adjacent functional layers in the device). Note that the atomic geometry of the perovskite surface layer remains flat in all AF-coated structures, meaning that the interaction between the perovskite surface atoms and the AF atoms does not cause noticeable lattice distortion which is very likely harmful to the materials stability. These values do not exhibit a clear trend, possibly because the LiF film has the highest density of atoms thus very likely the highest probability of Li-I and Pb-F bonding at the interface, while the K element has the smallest electronegativity so that the K-I bonds at the interface are stronger than Na-F and LiF — both can lead to stronger interaction between the perovskite and the AF layer. Another reason might be the largest strain (+1.7%, compared with the -0.55% for LiF and +0.73% for KF) with the NaF in the model system.

The formation energies of (neutral) $v_I$ are 1.90, 2.01, 2.35, and 2.44 eV for the bared (control), LiF-coated, NaF-coated, and KF-coated perovskite surface, respectively (as shown in **Figure 4B**). From them we can conclude that all ultrathin AF coatings can suppress the formation of $v_I$, which are believed to be one of the major sources of nonradiative recombination at the perovskite surfaces and the instability of materials and devices. The effectiveness of AF increases along with LiF < NaF < KF, agreeing well with the results of experimental measurements.

Considering the compactness and stability of the inorganic interlayers as the impenetrable (physically blocking) "walls", we expect the water and oxygen invasion into the perovskite layer and bidirectional ion migration at the perovskite-CTL interfaces to be largely prevented. To this end, we carried out time-of-flight secondary ion mass spectroscopy (ToF-SIMS) to probe the depth profiles of the atomic species within the sample (**Figure 4D**). The much shallower diffusion of $Ag^+$ into the perovskite layer, and the obviously reduced $I^-$-distribution in PTAA and especially $C_{60}$, evidence the physical-blocking functionality of the AF buffer layers. With the induced huge energy barrier for small particles to penetrate, it is also natural to expect that the AF buffer layers can well protect the perovskite surface against the interaction with environmental small molecules such as oxygen and water, which easily leads to material degradation**.** The binding energies of carbon from XPS results (**Figure 4C**) show C and C-N peaks located at 284.6 eV and 286.1 eV for all the perovskite samples. For KF modification, an additional peak at a binding energy of 292.9 eV emerges signifying the $K^+$ cations. Notably, the control film exhibits a strong signal at 288.1 eV which is attributed to the C=O bonding,[41] indicating a prevalent interaction with oxygen. As AF varies from LiF to NaF and KF, this undesired C=O peak gradually diminishes and finally almost disappears.

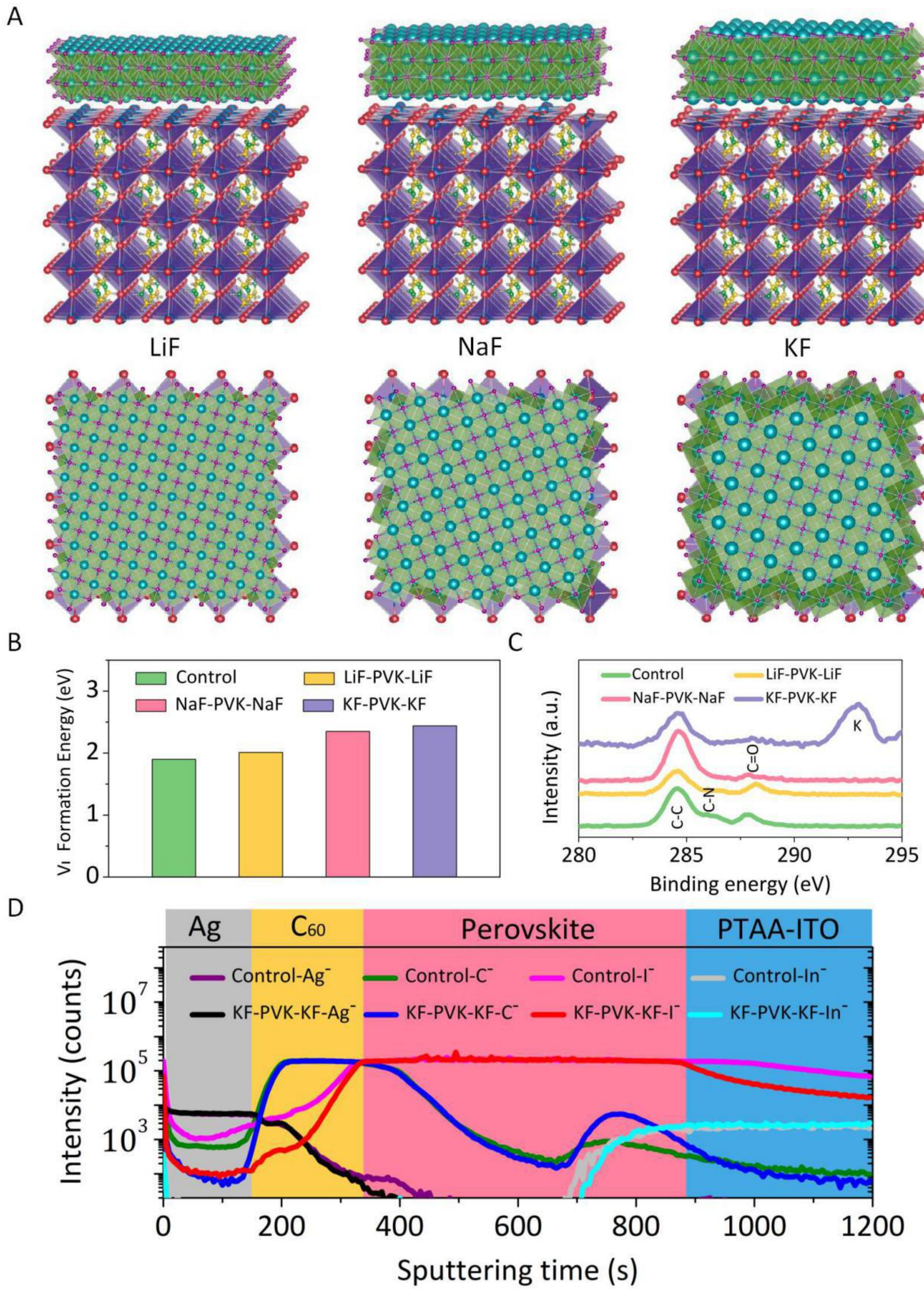

**Figure 4 (A)** DFT-optimized structures of $FAPbI_3$-AF complexes (left, middle, and right for A = Li, Na, and K, respectively; C, N, H, Pb, I, A, and F in green, yellow, gray, blue, red, turquoise, and purple, respectively). (B) Formation energy of surface $v_I$ defects calculated with DFT. (C) XPS spectra of the perovskite films without and with AF modification, showing the binding energy of C 1s bound to different elements (plus the K $2p_{3/2}$ of the KF-PVK-KF sample). (D) ToF-SIMS elemental depth profiles of control and KF-PVK-KF aged for 300 h of illumination under $N_2$ atmosphere.

## Performance of indoor−photovoltaic devices

We also investigated the potential of applying the inverted planar perovskite photovoltaic devices with inorganic bilateral buffer layers in indoor light-harvesting (schematic diagram shown in **Figure 5A**). As already reported, perovskite-based devices perform excellently under dim-light condition as a result of the large component-flexibility of materials and therewith the tunable bandgap.[9] **Figure 5B** shows the illumination spectrum and chromaticity coordinate of the white LED. This spectrum corresponds to a maximal PCE (detailed balance limit) of 53% associated with a bandgap of 1.9 eV. **Figure 5C** shows the indoor *J–V* curves of the devices with and without KF modification under a white LED source, with the parameters presented in **Table S5**. Under standard LED low illumination (1000 lux, color temperature = 5900 K, power density = 292 μW/cm$^2$), a PCE of 24.9% ($J_{SC}$ = 126 μA/cm$^2$, $V_{OC}$ = 0.81 V, FF = 71.2%) was achieved by the control device, while an outstanding performance of 35.7% PCE was obtained with the KF-modified device ($J_{SC}$ = 148 μA/cm$^2$, $V_{OC}$ = 0.93 V, FF = 75.7%). The performance improvement in indoor-photovoltaic applications upon the incorporation of KF interlayers is thus much more significant than in solar-light harvesting. We expect an even better performance of inverted planar perovskite devices with this inorganic bilateral strategy for indoor-light photovoltaics, for example, providing that the bandgap is tuned closer to the optimal value with proper materials engineering.

In **Figure 5D**, the relationship between $V_{OC}$ and the light intensity of different devices (solar spectrum) was measured, and the diode ideality factor $n$ ($1 < n < 2$) was calculated using the following equation:

$$V_{OC} \sim \frac{nKT}{e}\ln(\frac{I_{ph}}{I_0}) \qquad (1)$$

where $k$ is the Boltzmann constant, $T$ is the temperature, $e$ is the elementary charge, $I_{ph}$ is the photocurrent, and $I_0$ is the saturation current of the diode. By fitting the data, the ideality factors $n$ of the device are 1.35 and 1.12 for the control and the KF-modified devices, respectively. A larger $n$ value corresponds to a more severe trap-assisted SRH recombination. Therefore these results signify a successful suppression of defect-induced trap states with the help of the bilateral ultrathin KF interlayers.

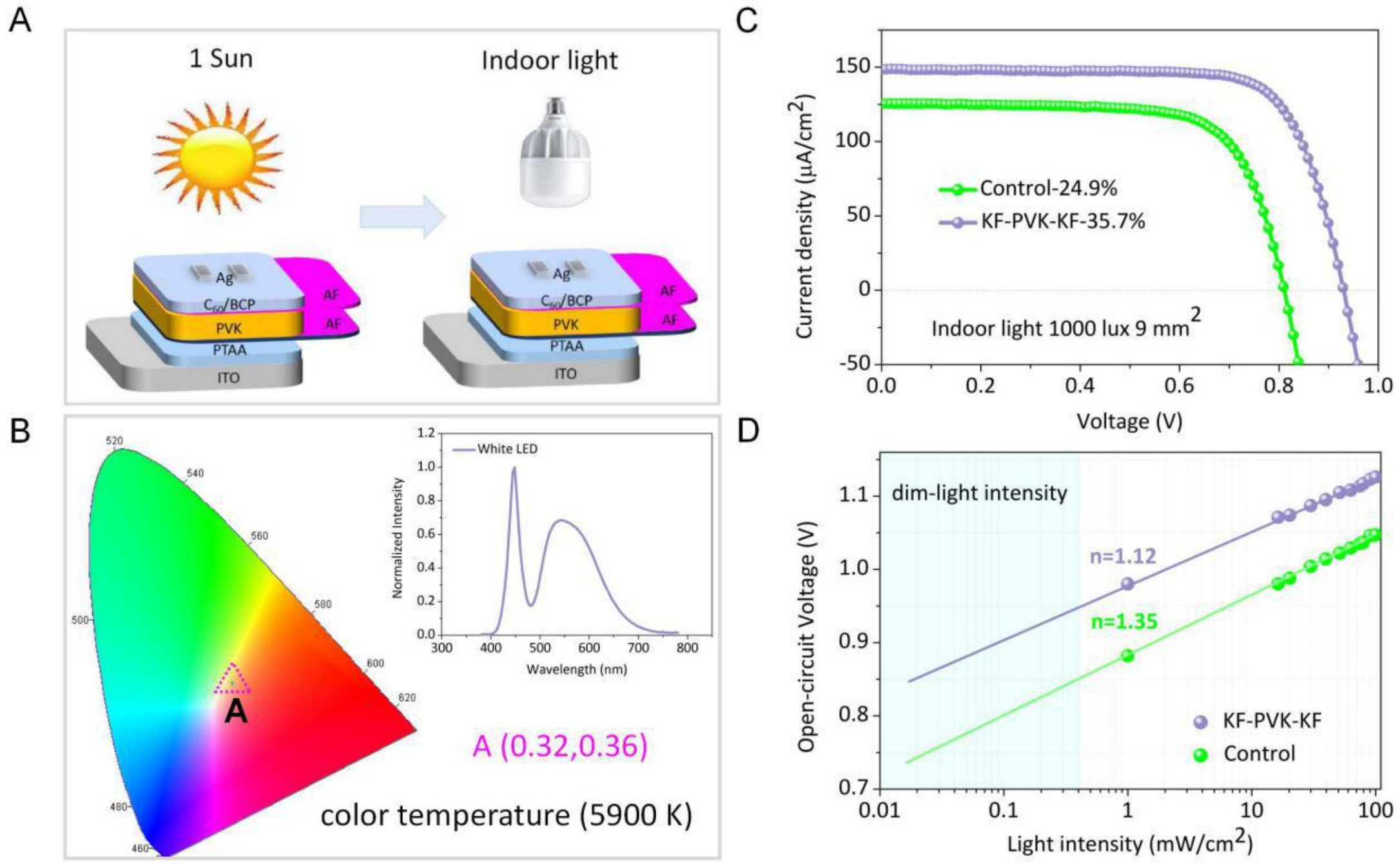

**Figure 5** (A) Schematic diagram of devices for control and KF-PVK-KF under one sun and indoor light. (B) Illumination spectra and chromaticity coordinate of white LED. (C) *J-V* characteristics of a 9 mm$^2$ device at 1000 lux. (D) $V_{OC}$ measured at different light intensity under AM 1.5G with extrapolation to the dim-light-intensity region.

# Conclusion

In summary, we have proposed an inorganic strategy of surface engineering to improve the efficiency and stability of PSCs. By coating ultrathin KF films at the surfaces of the triple-cation perovskite active layer with a nearly complete coverage, that is, modifying both perovskite-ETL and perovskite-HTL interfaces, the formation of defects at the perovskite surfaces and grain boundaries as well as the bidirectional ion migration at the interfaces were substantially suppressed as evidenced by both experimental and theoretical analysis. As a result, the defect-induced nonradiative recombination was significantly reduced thus leading to the more effective interfacial charge extraction and the less material degradation. With the optimized devices we obtained a PCE of 22.02% which is among the best records for inverted planar PSCs. The unencapsulated devices exhibit exceptional operational stability (500 h, ISOS-L-2) and thermal stability (1000 h, ISOS-D-2). A remarkable PCE of 35.7% was achieved under a standard white-light LED source with these devices, which also belongs to the best records of indoor photovoltaics to-date and shows a promising future in indoor-light harvesting. Our approach thus offers a new solution to the critical problem of today's perovskite optoelectronic community, and opens up space for further advancement of perovskite-based photovoltaic techniques toward high efficiency and high stability in both outdoor and indoor applications.

**Characterizations:**

A field emission scanning electron microscope (SEM) (Quanta 250, FEI, USA) was used to investigate the morphology and crystallinity. The crystalline structure on ITO substrate is performed by a X-ray diffractometer (Bruker D8 ADVANCE) with Cu Kα radiation. The absorption was obtained by ultraviolet-visible spectrophotometer (HITACHI U-3010, Japan). The D-XPS characterization was carried out by X-ray photoelectron spectroscopy (ESCALAB Xi+, Thermo Fisher Scientific). The photovoltaic performance was estimated under a AAA solar simulator (XES-301S, SAN-EI), AM 1.5G irradiation with an intensity of 100 mW/cm$^2$. The photocurrent density-voltage (*J-V*) curve was measured using a Keithley (2602 Series Sourcemeter) with scan rate 0.01 V/s within the range from -0.1 to 1.2 V. The area of each device, calibrated by the shadow mask, was 4.92 mm$^2$**.** The photoluminescence intensity and lifetime images of each perovskite sample were acquired simultaneously by the ISS (ISS Inc., Champagne, IL) PL1 FastFLIM system (details in supporting information).

KPFM Characterizations: The KPFM was used to study the versatile perovskite films that participate in the formation of different heterojunctions in nanometer resolution by collecting the contact potential differences under illumination. KPFM is a surface potential detection method that determines the CPD during scanning by compensating the electrostatic forces between the probe and the sample. KPFM was carried out with Asylum Research Cypher atomic force microscope using a silicon probe coated with Ti/Ir (ASYELEC-01-R2) with a force constant of 0.5 – 4.4 nN nm$^{-1}$ and a tip radius of 28 ± 10 nm. The Nap mode method was used, including two passes-the first pass is used to acquire the topographic height using tapping mode, the second one which is raised above the surface with a lift height of 40 nm and scanned to acquire the potential offset between the tip and the sample through a DC voltage feedback loop. KPFM measurements have shown consistent results for repetitive measurements. All measurements were performed under continuous light illumination at an angle about 45. Irradiation was provided by a light source with a light intensity of 20 mW/cm$^2$. The AFM laser has a wavelength of above 600 nm (E > 2.07 eV). The intensity of the laser is very low, so the laser has little influence on our measurements.

**Acknowledgement**

This work is financially supported by the National Natural Science Foundation of China (Grant Nos. 61604121, 61935016). China Postdoctoral Science Foundation (Grant No. 2019M663717, 2020T130502), Scientific Research Plan Projects of Shaanxi Education Department (Grant No. 17JK0700), Natural Science Basic Research Plan in Shaanxi Province of China (Grant No. 2019JQ-119), The authors thank Chunyan Lu and Yanyan Wang at Ningbo University for her help in using KPFM. The authors thank Dr. Yuansheng Sun and Dr. Dongsheng Wu at ISS, Inc. for his help of FastFLIM records. We also thank Dr. Liu at the Instrumental Analysis Center of Xi'an Jiaotong University for her assistance with XPS analysis. This work is also supported by Xi’an Jiaotong University’s HPC Platform.

**Author Contributions**

Jie Xu and Jun Xi conceived and designed the experiments, including fabrication and analysis of the films and devices. Jinbo Chen contributed to absorption spectra measurement and analysis. Jie Xu wrote the first version of the manuscript. Jun Xi contributed to the modification of manuscript. All authors discussed the results and contributed to the final version of the paper. Hua Dong, Jingrui Li and Zhaoxin Wu supervised the project.

**Notes:** The authors declare no competing financial interest.

**Appendix A. Supplemental Information**

**Supplementary data associated with this article can be found in the online version at.**

# Impermeable Inorganic "Walls" Sandwiching Photoactive Layer toward Inverted Perovskite Solar and Indoor Photovoltaic Devices

*Jie Xu,[1,#] Jun Xi,[2,#,*] Hua Dong,[1,7,*] Namyoung Ahn,[4] Zonglong Zhu,[5] Jinbo Chen,[1] Peizhou Li,[1] Xinyi zhu,[1] Jinfei Dai,[1] Ziyang Hu,[6] Bo Jiao,[1] Xun Hou,[1] Jingrui Li,[3,*] and Zhaoxin Wu[1,7*]*

[1]Key Laboratory for Physical Electronics and Devices of the Ministry of Education & Shaanxi Key Lab of Information Photonic Technique, School of Electronic Science and Engineering, Xi'an Jiaotong University, No.28, Xianning West Road, Xi'an, 710049, China.

[2]Zernike Institute for Advanced Materials, University of Groningen, Nijenborgh 4, 9747 AG Groningen, the Netherlands

[3]Electronic Materials Research Laboratory, Key Laboratory of the Ministry of Education & International Center for Dielectric Research, School of Electronic Science and Engineering, Xi'an Jiaotong University, Xi'an 710049, China.

[4]Chemistry Division, Los Alamos National Laboratory, Los Alamos, NM 87545, USA.

[5]Department of Chemistry, City University of Hong Kong, Kowloon, Hong Kong.

[6]Department of Microelectronic Science and Engineering, Ningbo University, Ningbo 315211, China.

[7]Collaborative Innovation Center of Extreme Optics, Shanxi University, Taiyuan 030006, China.

**[#]These two authors contributed equally to this work.**
***Corresponding authors.**
**E-mail:**
**j.xi@rug.nl**
**donghuaxjtu@mail.xjtu.edu.cn**
**jingrui.li@mail.xjtu.edu.cn**
**zhaoxinwu@mail.xjtu.edu.cn**

**Supplementary Materials for Materials and Methods**

**Materials and reagents**

Methylammonium iodide (MAI, 99.99%), formamidinium iodide (FAI, 99.5%), cesium iodide (CsI, 99.99%), and poly[bis(4-phenyl)(2,4,6-trimethylphenyl)amine] (PTAA, Mn < 6000) were purchased from Xi'an Polymer Light Technology Corp (China). Lead iodide ($PbI_2$, > 99.99%) was purchased from TCI (Japan). 2,3,5,6-Tetrafluoro-7,7,8,8-tetracyanoquinodimethane (F4TCNQ, 99.0%), bathocuproine (BCP, 99.0%), and buckminsterfullerene ($C_{60}$, 99%) were purchased from Nichem (Taiwan). The interface materials, lithium fluoride (LiF), sodium fluoride (NaF), and potassium fluoride (KF), were purchased from VIZUCHEM (China) and used without purification. Besides, some reagents, including chlorobenzene (CB, Extra Dry, 99.8%), N,N-dimethylformamide (DMF, anhydrous, 99.8%), and dimethyl sulfoxide (DMSO, anhydrous, 99.8+%) were purchased from commercial sources (Acros) and used as received. Metal material silver was obtained from commercial sources (China New Metal Materials Technology Co. Ltd.) with high purity (99.99%).

**Perovskite precursor solution**

A mixed solution of FAI (1.14 M), $PbI_2$ (1.4 M), and MAI (0.15 M) was dissolved in N,N-dimethylformamide (DMF) and dimethyl sulfoxide (DMSO) (4:1, v/v) with an excess amount of $PbI_2$. 45.6 μL CsI solution (1.5 M) was added into the mixed solution for the desired composition.

**Doped-PTAA solution**

We first obtained a solution of F4TCNQ in CB (1.5 mg/ml), which was stirred at 70 °C for 15 min. F4TCNQ solution was then added into PTAA solution (1.5 mg/ml in CB) with a weight ratio of 1%.

**Hole-transport layer**

The indium tin oxide (ITO) patterned glass substrates with sheet resistance of about 15 Ω/square were cleaned with deionized water and organic solvents, and then exposed to UV–ozone ambience for 15 min. The substrate's size is 2.5 × 2.5 $cm^2$. Then, the ITO substrates were transferred to a $N_2$-filled glovebox with $H_2O$ and $O_2$ concentrations of < 0.1 ppm. F4TCNQ-doped PTAA solution was spin-coated onto the ITO substrates at 6000 rpm (with a ramping rate of 2000 rpm/s), and the substrates were heated at 100 °C for 10 min in a $N_2$-filled glovebox. Then, the samples were transferred to a vacuum chamber without being exposed to air. LiF, NaF, or KF ultrathin film (3-nm thick) was then deposited via thermal evaporation in a vacuum chamber with the base pressure of < $4.5\times10^{-4}$ Pa. The ultrathin interlayer was pre-treated for perovskite films deposition.

**Perovskite films deposition**

The organic-inorganic lead halide perovskite solution was spin-coated on the PTAA-coated ITO samples by a two-step sequential process at 2000 rpm for 10 s with a ramping rate of 2000 rpm/s, and 6000 rpm for 30 s with a ramping rate of 6000 rpm/s. During the high-speed step, 250 μL of anhydrous CB was poured on the center of the spinning substrates 16 s prior to the end of the whole spinning process. Then, the films were quickly annealed at 100 °C for 60 min.

**Electron-transport layer and metal electrodes**

The samples were transferred to a vacuum chamber without being exposed to air. LiF, NaF, or KF ultrathin film (3-nm thick) was then deposited via thermal evaporation. The ultrathin interface layer was pre-treated to electron- transport layer deposition. $C_{60}$ (30 nm)/BCP (6 nm) were then evaporated. Subsequently, the metal electrode Ag was thermally evaporated with 120-nm thickness. All of the thermal evaporation rate ranged from 0.1 to 2 Å/s depending on different materials used. And all of the thermally evaporated steps were in a vacuum chamber with the base pressure of < $4.5\times10^{-4}$ Pa.

**t-DOS:**

Impedance spectroscopy was measured with an Electrochemical Analyzer Meter (CHI 660d). The thermal admittance spectroscopy (TAS) measurement was carried out without bias-voltage in dark. The trap density of states was derived from the angle frequency-dependent capacitance using the equation: The energetic profile of trap density of states (tDOS) can be derived from the angular frequency-dependent capacitance using the equation:

$$N_T(E_\omega) = -\frac{V_{bi}}{qW}\frac{dC}{d\omega}\frac{\omega}{k_B T}$$

where $C$ is the capacitance, ω is the angular frequency, $q$ is the elementary charge, $k_B$ is the Boltzmann's constant, and $T$ is the absolute temperature. $V_{bi}$ and $W$, extracted from capacitance–voltage measurements, are the built-in potential and depletion width, respectively. The applied angular frequency ω defines an energetic demarcation,

$$E_\omega = -k_B T \ln\left(\frac{\omega_0}{\omega}\right),$$

where $\omega_0$ is the attempt-to-escape frequency. R(CR) equivalent circuit for fitting the impedance spectra:

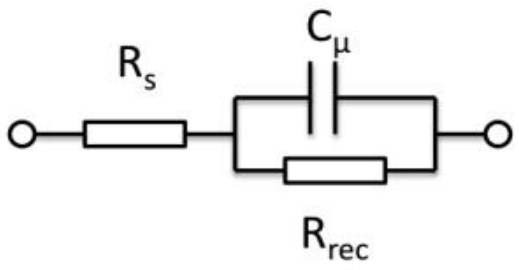

The trap states above the energy demarcation cannot contribute to the capacitance due to their slow rate to capture or emit charges.

The photoluminescence intensity and lifetime images of each Perovskite sample were acquired simultaneously by the ISS PL1 confocal time-resolved microscopy imaging system (www.iss.com). It is coupled to a Nikon Ti2U microscope equipped with a Nikon 40X / 0.95NA objective lens. The 488-nm pulsed diode laser with the tunable repetition rates from 1 Hz to 80 MHz was used as the excitation source. The laser repetition rate was set to be 500 KHz or 200 KHz for measuring samples. The emission light after a 647-nm long-pass filter (Semrock BLP01-647R) was collected by a single photon counting module avalanche photodiode (Excelitas SPCM-AQRH-45). For each sample, a field of 30 µm x 30 µm was scanned using the linear encoded automatic XY stage (ASI MS2000) at the sampling rate of 256 x 256 pixels. Both the laser and the stage scan were synchronized to the ISS FastFLIM data acquisition unit [1-4], to record the time-resolved data at each pixel of the image.

**Density-functional theory (DFT) calculations:**

In our DFT calculations, we used the Perdew−Burke−Ernzerhof exchange-correlation functional for solids (PBEsol)[1] in combination with the zero-order regular approximation (ZORA)[2] for the scalar relativistic effects. It was reported that the PBEsol functional can give proper description of atomic structures for halide perovskites especially the lattice constants.[3] Test calculations have also verified that PBEsol is a correct choice for alkali-fluoride materials.

For simplicity we used $FAPbI_3$ for the perovskite material, with neither considering the minor A-site cations Cs and MA, nor the possible occupancy of Li, Na, or K at the interstitial sites of the perovskite lattice. Based on the pre-optimized cubic structure of $FAPbI_3$ (2×2×2 supercell with single-cell lattice constant optimized at 6.33 Å), a slab model containing 5 layers of $PbI_2$ and 4 layers of FAI was constructed, thus mimicking a $PbI_2$-terminated surface model in consistent with the experimental condition of $PbI_2$-rich in the precursor. 4 layers of LiF, NaF, and KF (optimized rock-salt lattice constants 4.02 Å, 4.62 Å, and 5.30 Å, respectively) were placed above the optimized surface of $FAPbI_3$ to establish the respective AF-coated perovskite model systems. Since all lattice-constant ratios (for example, 6.33/4.02 = 1.57, $\sqrt{2}\cdot 6.33/5.30 = 1.69$) deviate from integers, we used large surface models (4×4, with lattice constant equal to 25.30 Å) and rotated the AF layer (around [001]) for a proper lattice match. As a result, 320 LiF (strain = -0.55%), 232 NaF (strain = +1.7%), or 180 KF (strain = +0.73%) units were deposited at the $FAPbI_3$ surface to form a 4-layer AF film. Finally, model systems for surface-iodine vacancies were constructed by removing one neutral I atom at the $FAPbI_3$ surface. The large model systems minimize the interaction between the $v_I$ point defects in neighboring unit cells. Nevertheless, we must pay the price of high computational costs.

All slab-model calculations were performed with fixing the atomic positions of the lowest two $PbI_2$ layers and two FAI layers, thus mimicking the bulk phase of perovskite. All other atoms including the alkali and fluoride ions were relaxed. A 40-Å thick vacuum layer was included to avoid the interaction between neighboring slabs. In addition, surface-dipole correction was considered in all slab-model calculations. All DFT calculations in this work were performed with the all-electron numeric-atom-centered orbital code FHI-aims[4,5]. The bulk structure of $FAPbI_3$ was calculated using the tier-2 basis set with a Γ-centered 6×6×4 *k*-point mesh. Tier-1 basis set a Γ-centered 2×2×1 *k*-point mesh were used for all slab-model calculations considering the large model systems.

**Table S1** Photovoltaic parameters for the inverted planar devices with different thickness of bilateral inorganic AF (A = Li, Na, K) interlayers.

| | $J_{SC}$ (mA/cm$^2$) | $V_{OC}$ (V) | FF (%) | PCE (%) |
|---|---|---|---|---|
| LiF-1 nm | 22.67 | 1.06 | 80.0 | 19.22 |
| LiF-2 nm | 22.65 | 1.085 | 80.0 | 19.66 |
| LiF-3 nm | 22.53 | 1.095 | 80.1 | 19.76 |
| LiF-5 nm | 17.47 | 1.086 | 75.3 | 14.29 |
| LiF-7 nm | 17.43 | 1.083 | 74.5 | 14.06 |
| NaF-1 nm | 22.69 | 1.06 | 79.8 | 19.19 |
| NaF-2nm | 22.66 | 1.09 | 79.9 | 19.73 |
| NaF-3 nm | 22.57 | 1.12 | 79.9 | 20.20 |
| NaF-5 nm | 17.53 | 1.10 | 75.7 | 14.60 |
| NaF-7 nm | 17.49 | 1.09 | 75.1 | 14.32 |
| KF-1 nm | 23.10 | 1.10 | 80.3 | 20.40 |
| KF-2 nm | 22.98 | 1.11 | 80.8 | 20.61 |
| KF-3 nm | 22.77 | 1.15 | 81.3 | 21.29 |
| KF-5 nm | 17.89 | 1.148 | 76.1 | 15.63 |
| KF-7 nm | 17.50 | 1.142 | 75.3 | 15.05 |

**Table S2** Photovoltaic parameters of champion devices.

| | $J_{SC}$ (mA/cm$^2$) | $V_{OC}$ (V) | FF (%) | PCE (%) |
|---|---|---|---|---|
| Control | 22.56 | 1.067 | 79.8 | 19.20 |
| LiF-3 nm | 22.76 | 1.100 | 80.3 | 20.10 |
| NaF-3 nm | 22.77 | 1.127 | 80.0 | 20.53 |
| KF-3 nm | 23.40 | 1.152 | 81.7 | 22.02 |

**Table S3** Summary of performance parameters of reported inverted PSCs modified by interlayers.

| Structure | $PCE_{max}$ (%) | $E_g$ (V) | $V_{OC}$-control (V) | $V_{OC}$-modified (V) | $\Delta V_{OC}$ (V) | $V_{OC}$-loss (V) | Ref. |
|---|---|---|---|---|---|---|---|
| ITO/PTAA/PMMA/Perovskite/C60/BCP/Ag | 20.57 | 1.64 | 1.08 | 1.12 | 0.04 | 0.52 | 1 |
| ITO/NiOX/KCl/CsFAMA/PCBM-ZrAcac-Ag | 20.96 | 1.60 | 1.07 | 1.15 | 0.08 | 0.45 | 2 |
| ITO/PTAA/$MAPbI_3$/$PC_{60}BM$/CMB-vTA/AZO/Ag | 17.15 | 1.61 | 1.077 | 1.090 | 0.013 | 0.52 | 3 |
| ITO/PEDOT:PSS/PEDOT:PSS-$VO_X$/MAPbI3/PCBM/$C_{60}$/LiF/Al | 18.0 | 1.61 | 0.84 | 1.02 | 0.18 | 0.59 | 4 |
| ITO/PTAA/MAPbI3/PCBB-3N-3I/PCBM/Al | 21.10 | 1.6 | 1.068 | 1.105 | 0.037 | 0.495 | 5 |
| Glass&TCO/P-TPD/PFN/Perovskite/I-PFC10(SAM)/$C_{60}$/BCP/Ag | 21.3 | 1.62 | < 1.1 | 1.18 | 0.08 | 0.44 | 6 |
| ITO/PTAA/$Cs_{0.05}(FA_{0.85}MA_{0.15})_{0.95}Pb(I_{0.85}Br_{0.15})_3$/PC61BM/ZrL3:bis-C60/Ag | 22.02 | 1.6 | 1.17 | 1.20 | 0.03 | 0.4 | 7 |
| ITO/PTAA/KF/$Cs_{0.05}(FA_{0.88}MA_{0.12})_{0.95}Pb(I_{0.88}Br_{0.12})_3$/KF/$C_{60}$/BCP/Ag | 22.02 | 1.55 | 1.067 | 1.152 | 0.085 | 0.398 | This work |

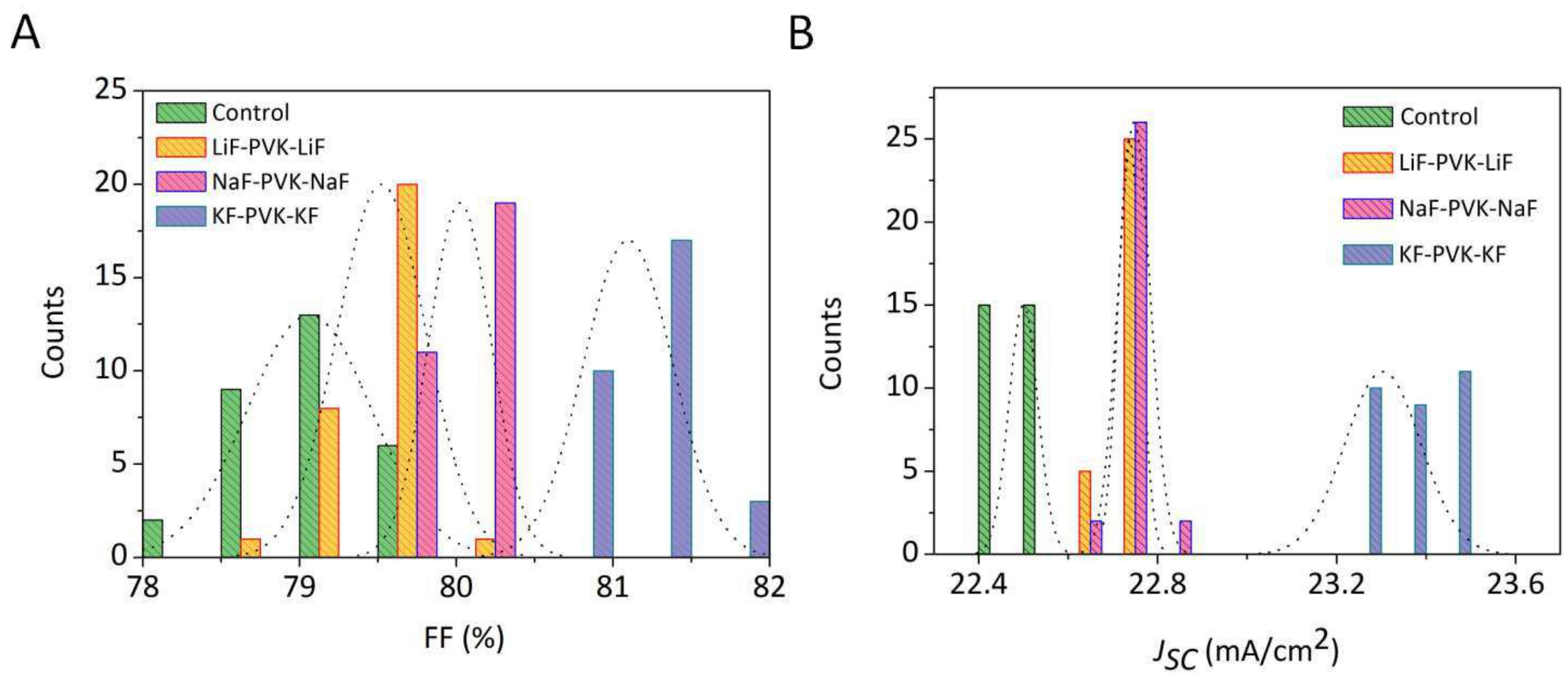

**Figure S1** Statistical distribution of (A) FF and (B) $J_{SC}$ for 120 devices (30 per structure). PVK stands for perovskite.

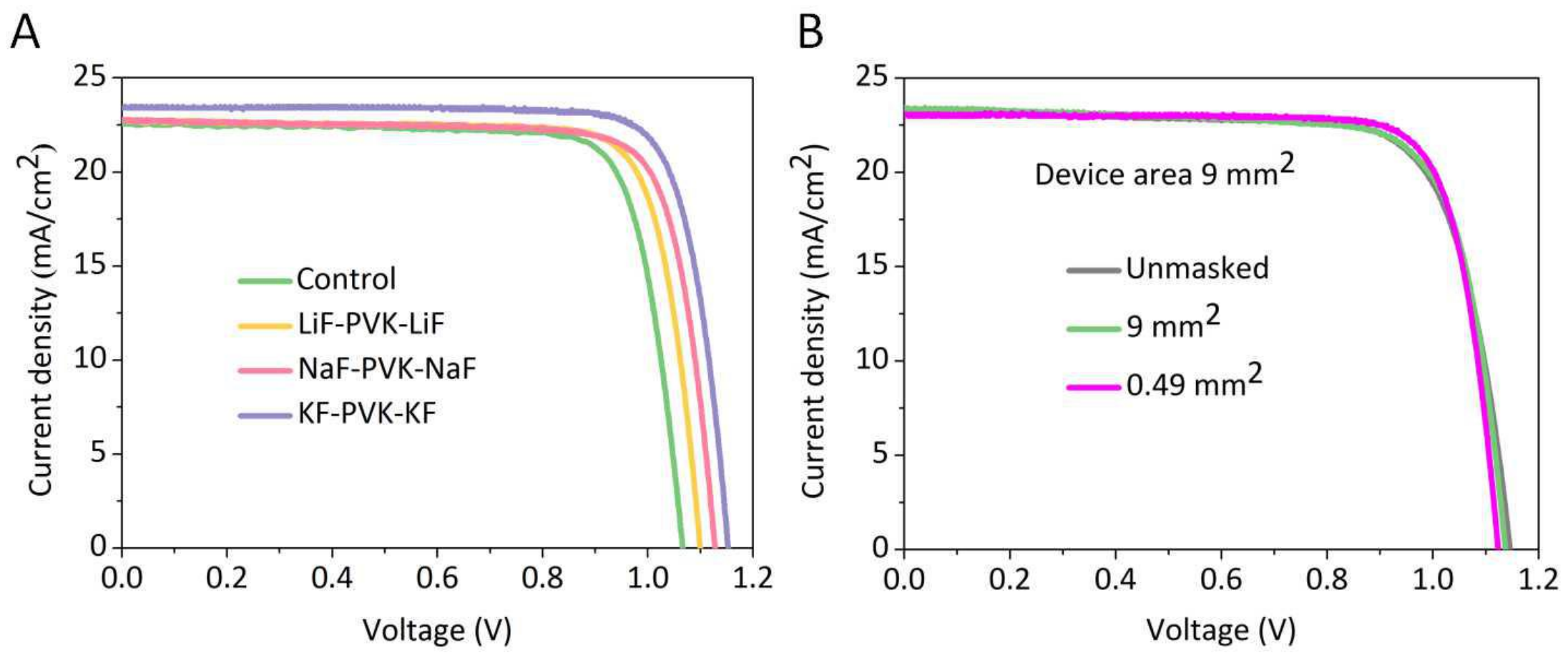

**Figure S2** (A) *J-V* curves of champion devices. (B) *J-V* curves of perovskite cell with a device active area of 9 mm$^2$ measured with two different mask apertures (one being the active area, while another are smaller than the active area and densities are determined by dividing the current by the smaller of the areas).

中 国 计 量 科 学 研 究 院

## Appendix: Summary of Certificate

**NIM Certificate No.: GXtc2019-1359**
**Client: Xi`an Jiaotong University**
**DUT S/N：28#-5-M1-F-01**
**Date of Test: 06/06/2019**
**Manufacturer：Xi`an Jiaotong University**
**Type: Perovskite Solar Cell**
**Temperature Sensor/Control System: None**
**Mask: An aperture area of 4.912 mm² (Certificate No.: CDjc2019-0206)**
**Environmental conditions at the time of calibration: (24.6±1) ℃, RH (60.0±2) %**

The calibration has been conducted by the PV Metrology Lab of NIM (National Institute of Metrology, China). Measurement of irradiance intensity and all other measurements are traceable to the International System of Units (SI). The performance parameters reported in this certificate apply only at the time of the test for the sample.

| Area ($mm^2$) | $I_{sc}$(mA) | $V_{oc}$(V) | $P_{max}$(mW) |
|---|---|---|---|
| 4.912 | 1.141 | 1.094 | 1.000 |
| $I_{max}$(mA) | $V_{max}$(V) | FF (%) | $\eta$ (%) |
| 1.076 | 0.930 | 80.2 | 20.4 |

**I-V Characterization Methods:**
Refer to IEC60904-1 2017: Measurement of photovoltaic current-voltage characteristics
According to JJF 1622-2017: Calibration Specification of Solar Cells：Photoelectric Properties

**Secondary Reference Cell:**
Device S/N: 81#
Device Material: Mono-Si

**Solar Simulator:**
Classification: AAA (Double-light source: Xeon and Halogen)
Total irradiance: 1000 W/m² based on $I_{sc}$ of the above Secondary Reference Cell.

| Issue Date | 06/06/2019 |
|---|---|

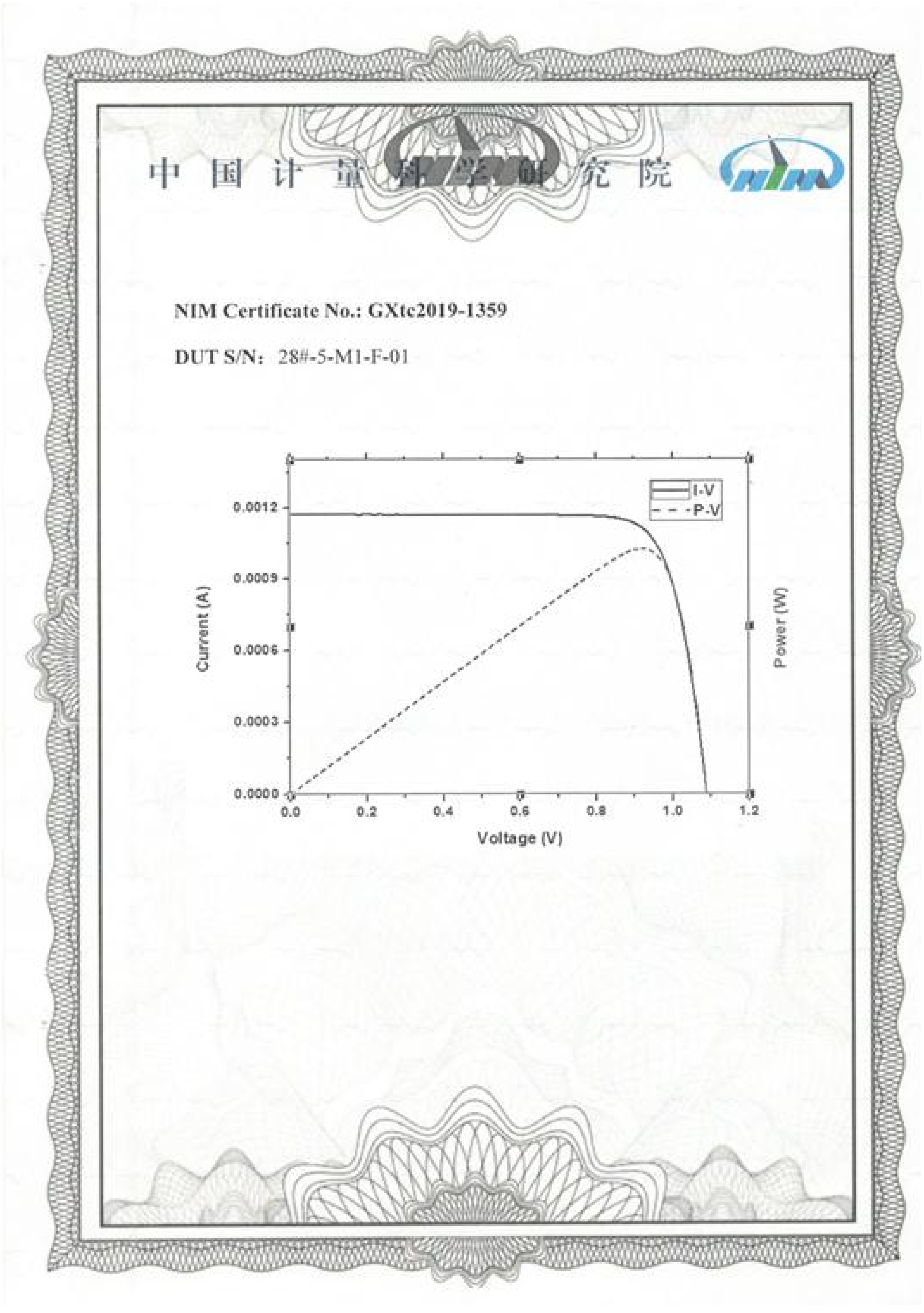
中国计量科学研究院

NIM Certificate No.: GXtc2019-1359

DUT S/N：28#-5-M1-F-01

**Figure S3** PCE testing report of an inverted planar heterojunction PSC by an accredited PV Metrology Laboratory of NIM (National Institute of Metrology, China) verified a PCE of 20.4% ($J_{SC}$ = 23.22 mA/cm$^2$ , $V_{OC}$ = 1.094 V, and FF = 80.2%) with negligible hysteresis. The cell was tested in air without encapsulation or protection during the test process.

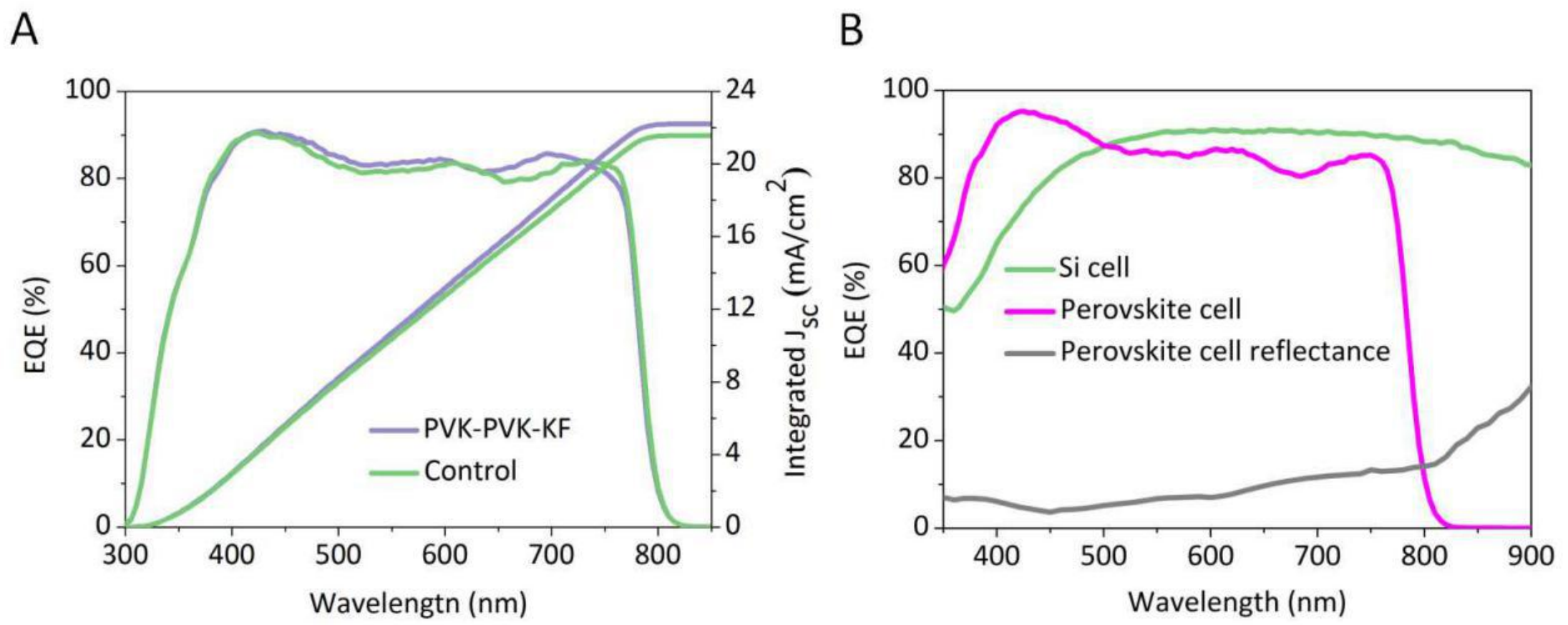

**Figure S4** (A) EQE spectra and integrated short-circuit current density of the control and the bilaterally KF-modified devices. (B) IPCE spectra for Si cell and perovskite cells, and total reflectance of perovskite cell illuminated from front side.

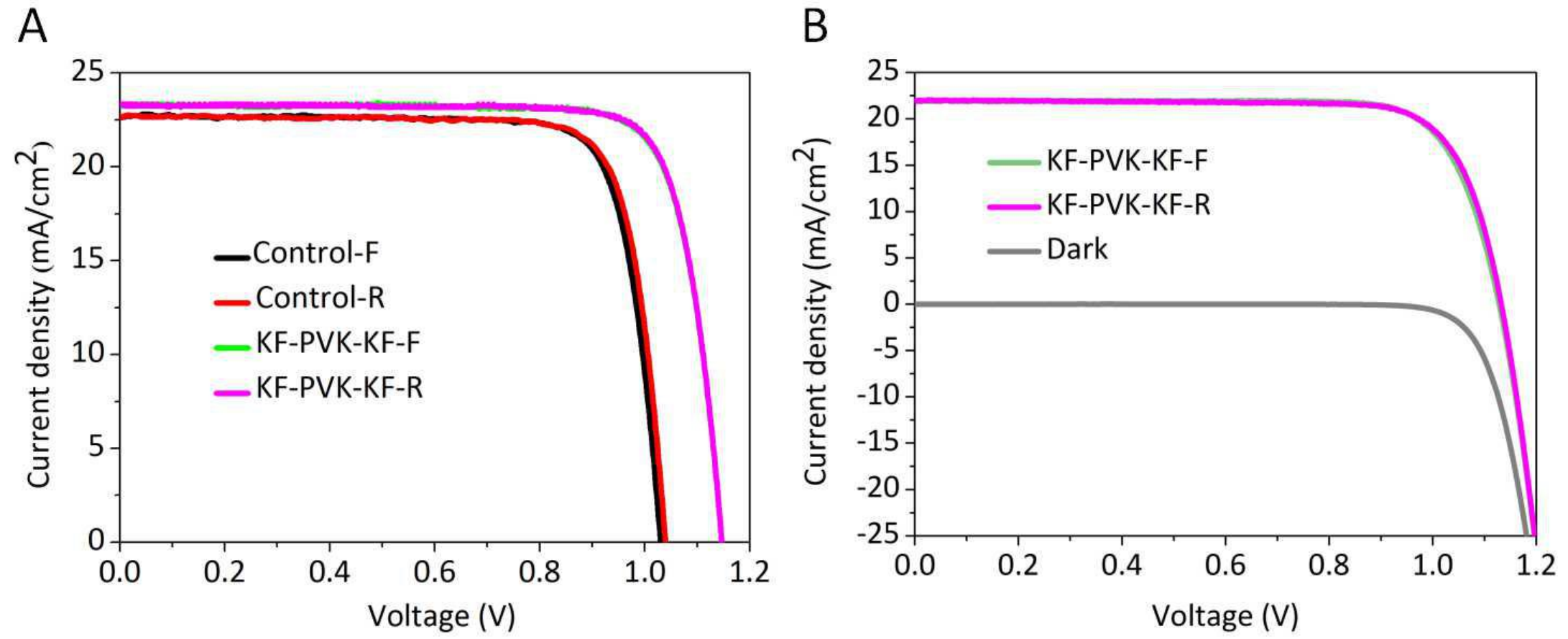

**Figure S5** *J-V* characteristic from forward scan (F, from $J_{SC}$ to $V_{OC}$) and reverse scan (R, from $V_{OC}$ to $J_{SC}$) determined under simulated AM 1.5G illumination for the control and the KF-perovskite-KF device with 0.49 mm$^2$ (A), and the KF-perovskite-KF device with1 cm$^2$ solar cells (B).

**Table S4** Photovoltaic parameters from forward and reverse scan determined under simulated AM 1.5G illumination for the control and the KF-perovskite-KF devices.

| | $J_{SC}$ (mA/cm$^2$) | $V_{OC}$ (V) | FF (%) | PCE (%) | HI (%) |
|---|---|---|---|---|---|
| Control-R | 22.43 | 1.065 | 80.1 | 19.15 | 3.3 |
| Control-F | 22.55 | 1.048 | 78.4 | 18.53 | |
| KF-3 nm-R | 23.26 | 1.146 | 81.8 | 21.80 | 0.3 |
| KF-3 nm-F | 23.34 | 1.146 | 81.2 | 21.73 | |

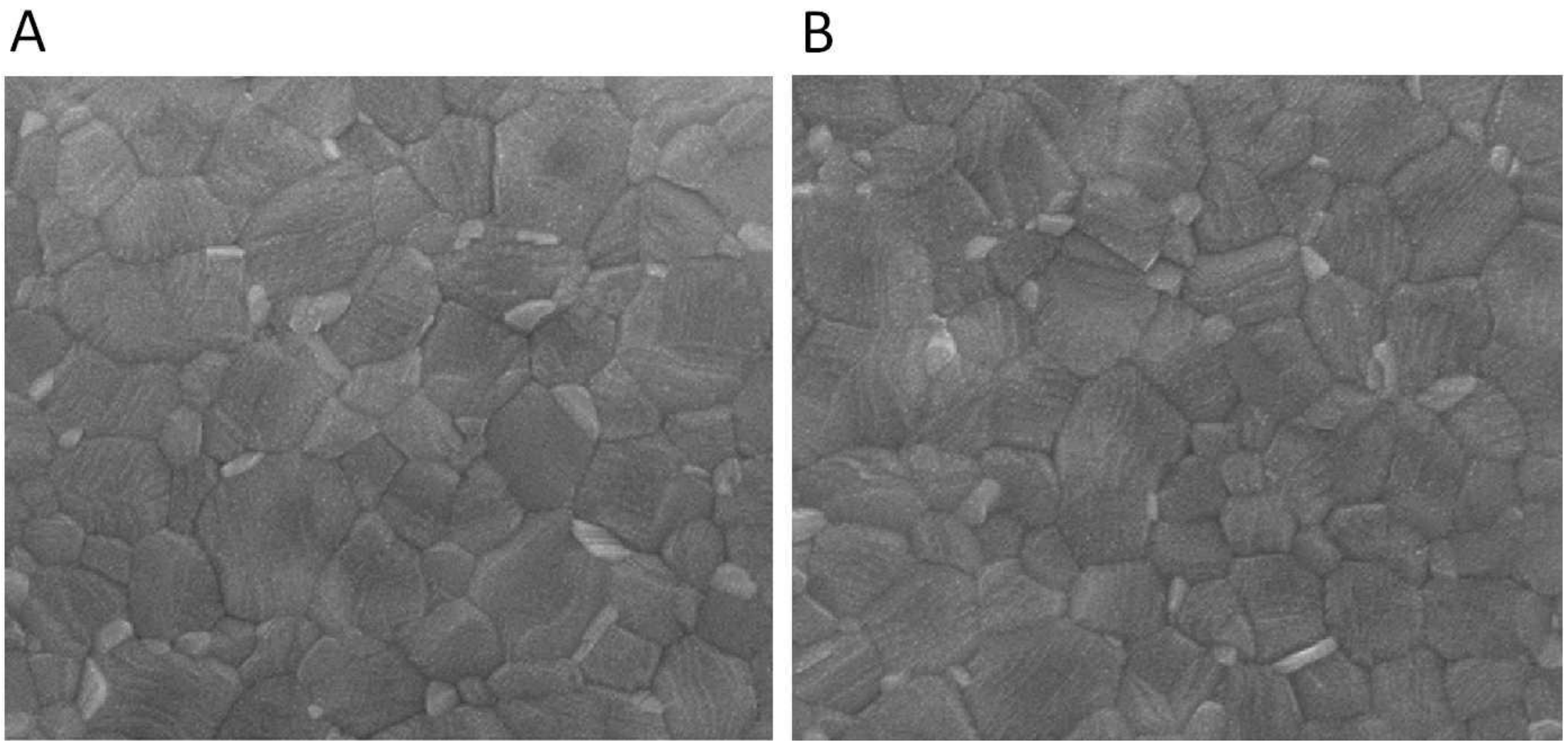

**Figure S6** SEM images of perovskite films modified with bilateral inorganic "walls": (A) LiF and (B) NaF.

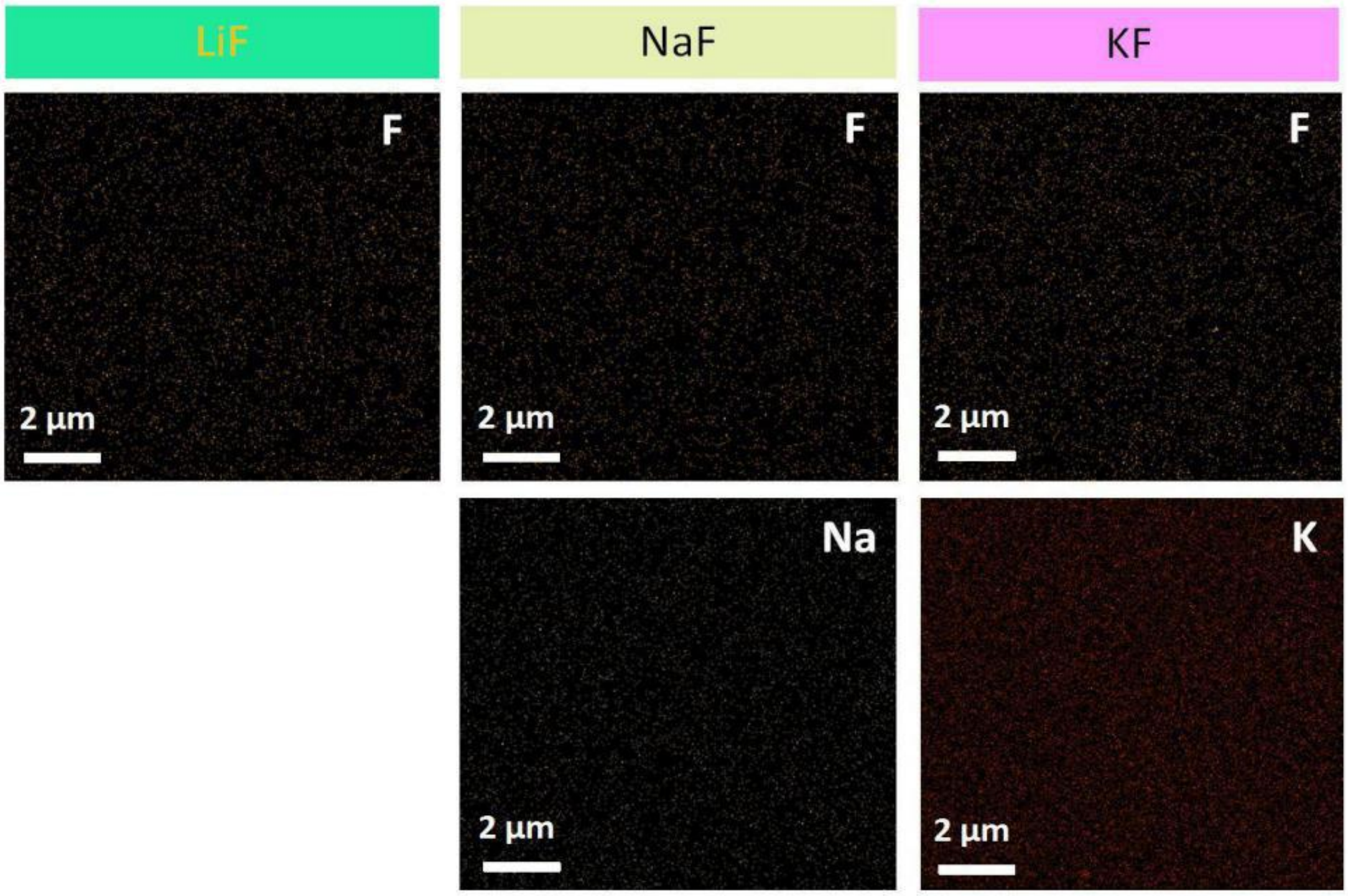

**Figure S7** Elemental maps of F, Na, and K on perovskite samples modified with bilateral inorganic (LiF, NaF, and KF) "walls" from energy dispersive X-ray (EDX) spectroscopy.

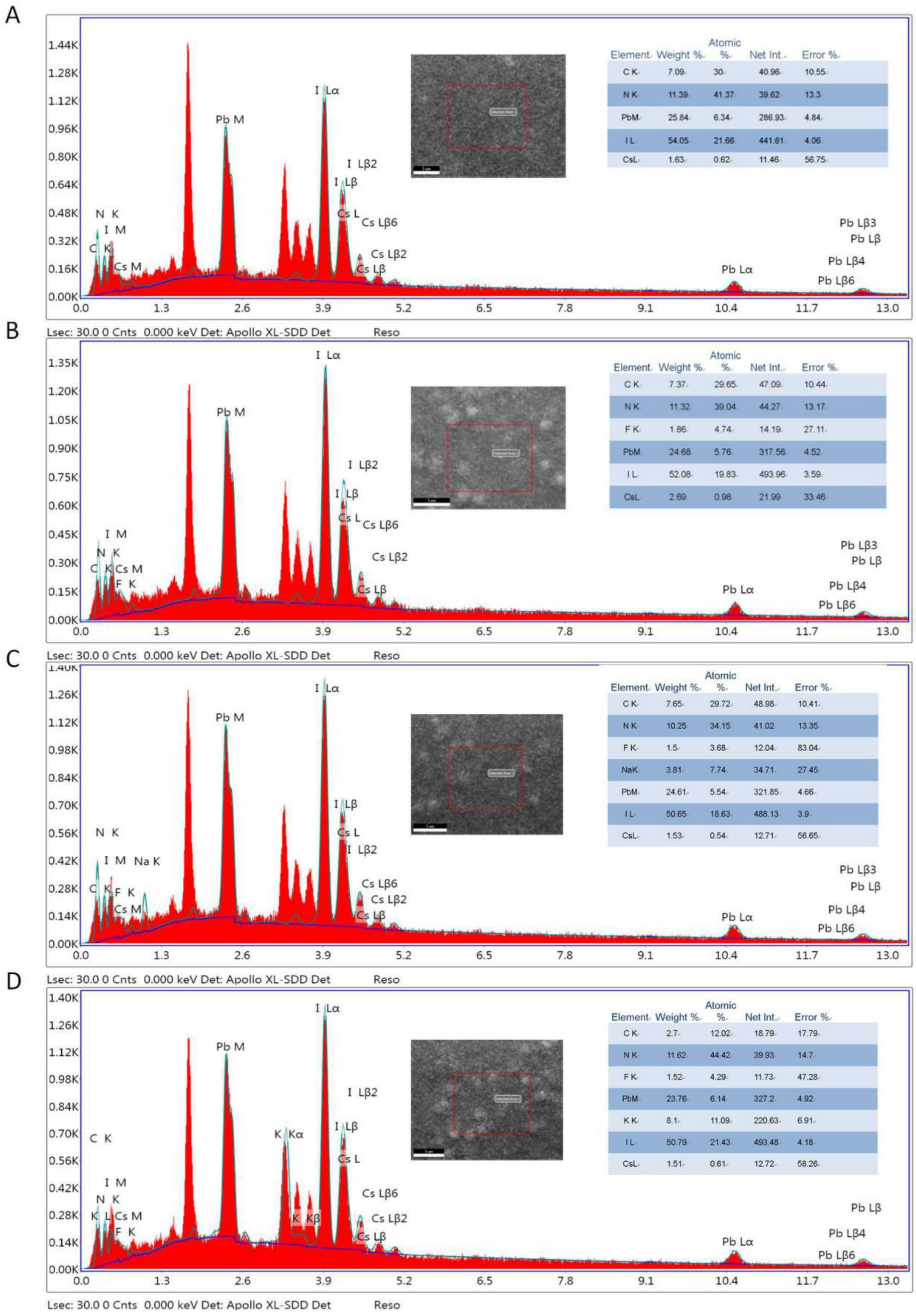

**Figure S8** SEM image of $Cs_{0.05}(FA_{0.88}MA_{0.12})_{0.95}PbI_3$ grains, EDX spectra at four spots randomly selected on the perovskite grains, and element ratio obtained from the corresponding EDX spectra: (A) control, (B) LiF-perovskite-LiF, (C) NaF-perovskite-NaF, and (D) KF-perovskite-KF.

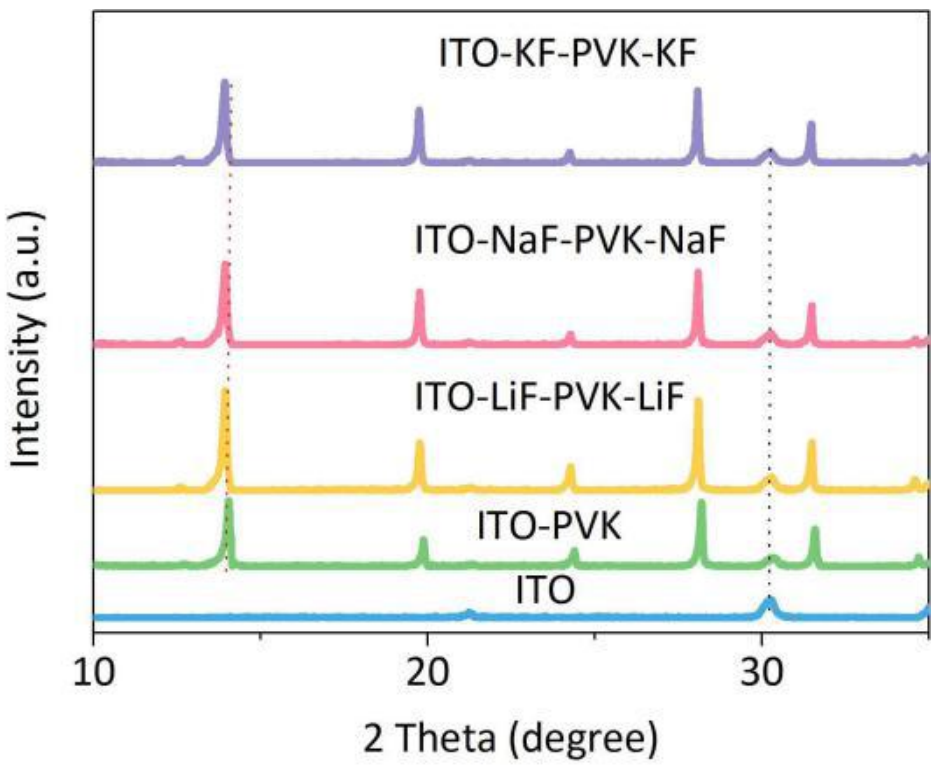

**Figure S9** XRD patterns of the control and bilaterally AF-modified perovskite films.

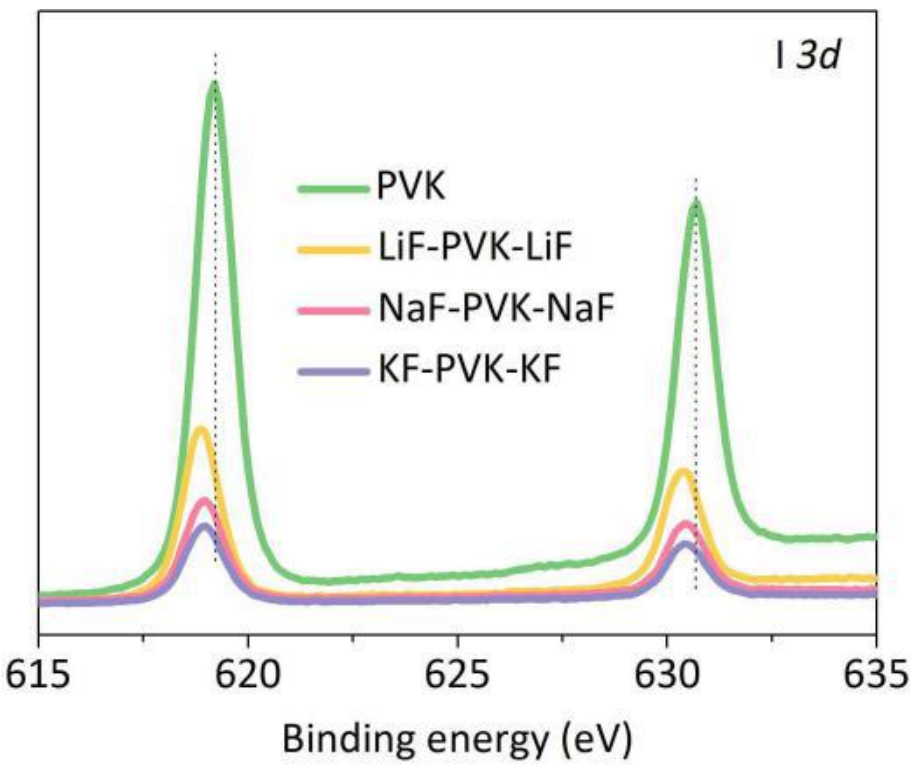

**Figure S10** XPS of the I 3d core-level of all four investigated samples.

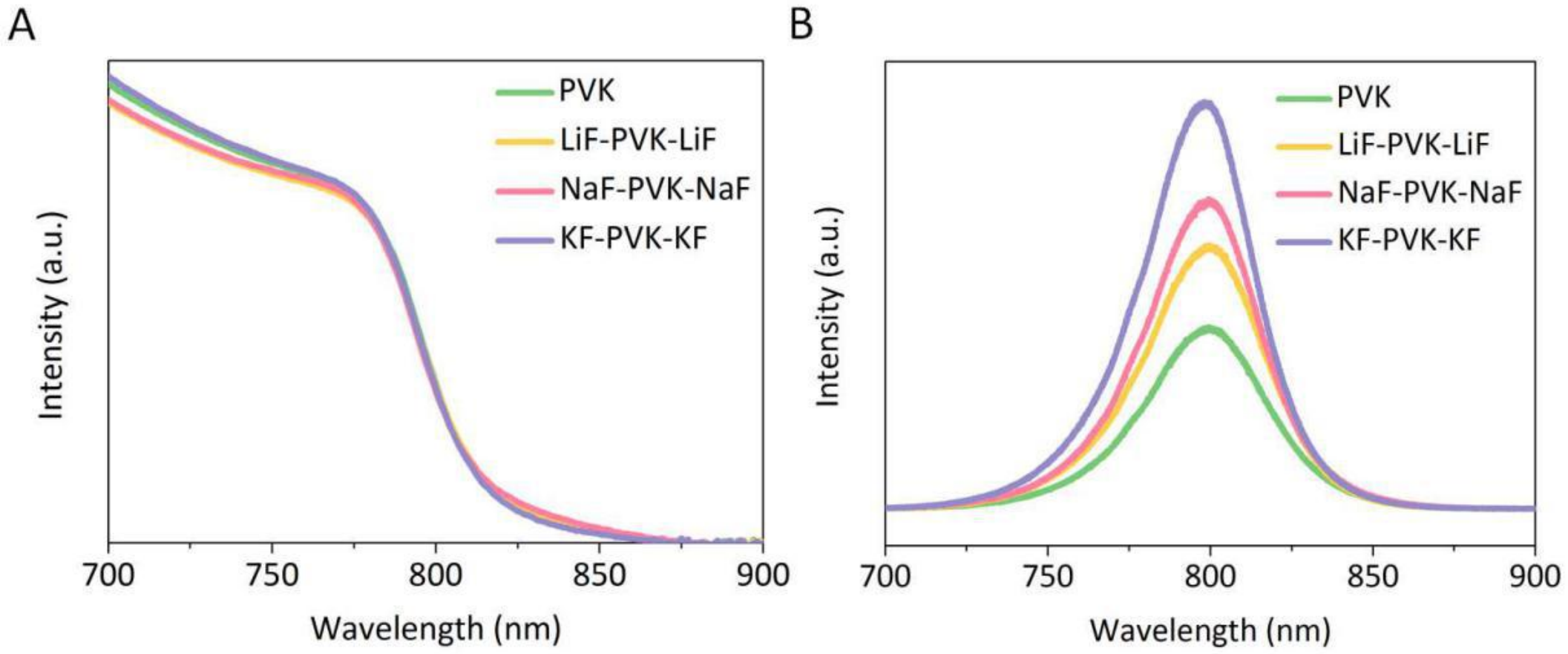

**Figure S11** (A) Absorption and (B) PL spectrum of all four investigated samples.

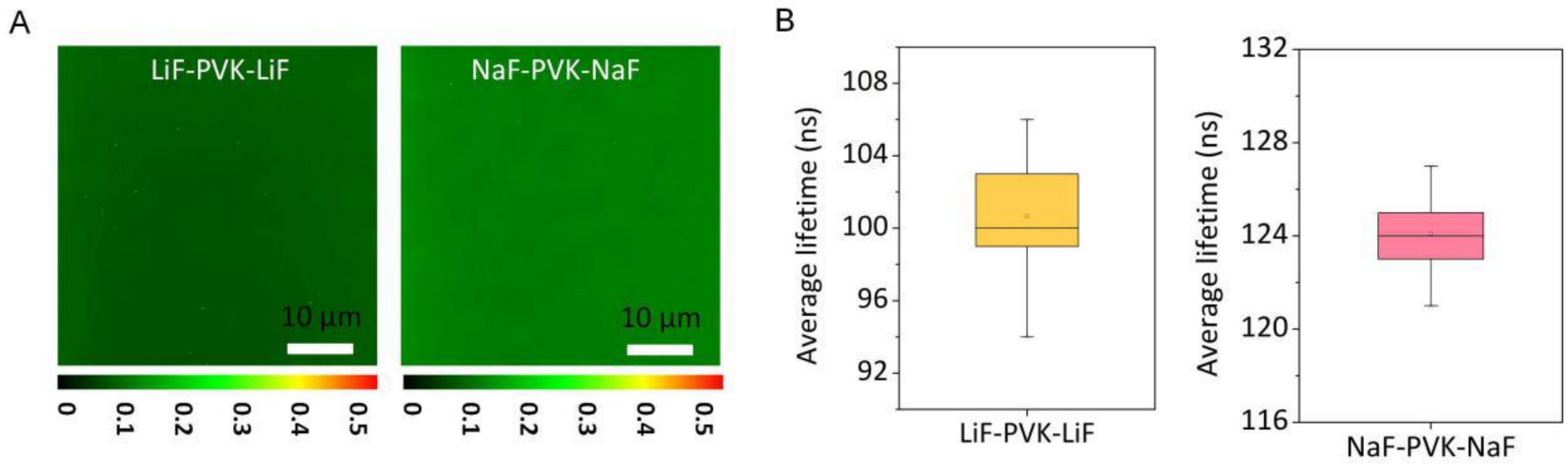

**Figure S12** (A) TRPL-mapping (scanning: 50 μm × 50 μm) and (B) statistics and average lifetimes of TRPL-mapping for the LiF-perovskite-LiF and the NaF-perovskite-NaF films.

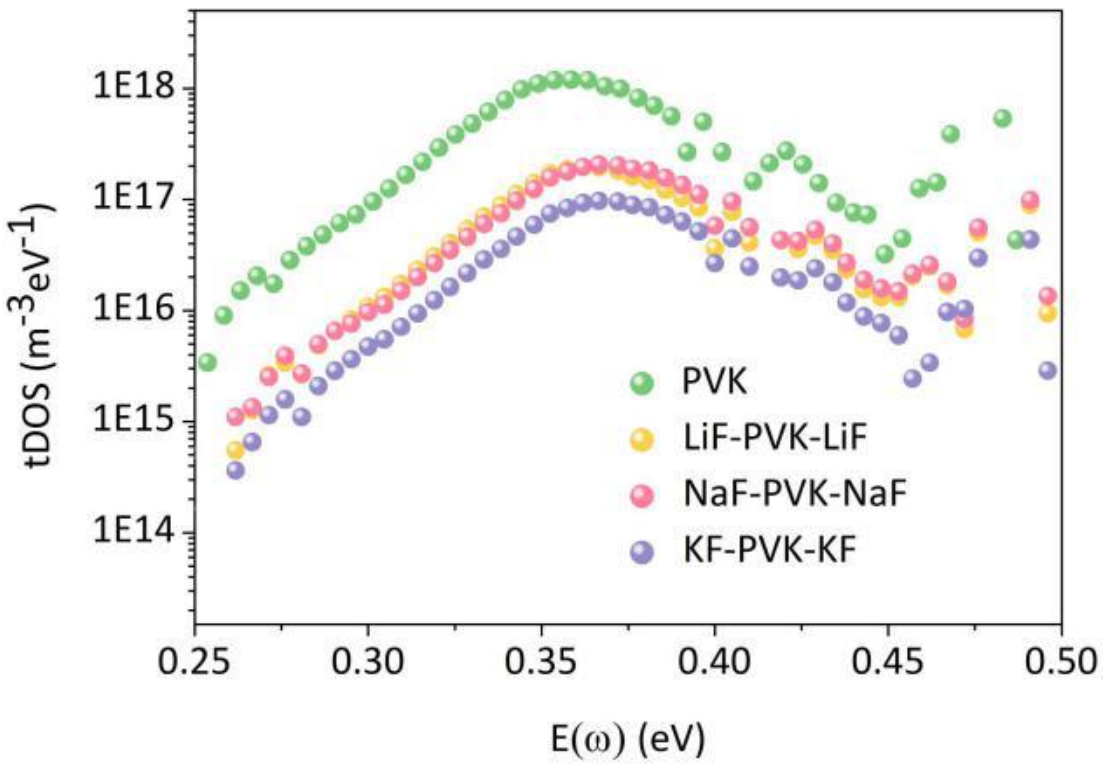

**Figure S13** Trap density of states (t-DOS) of all four investigated samples.

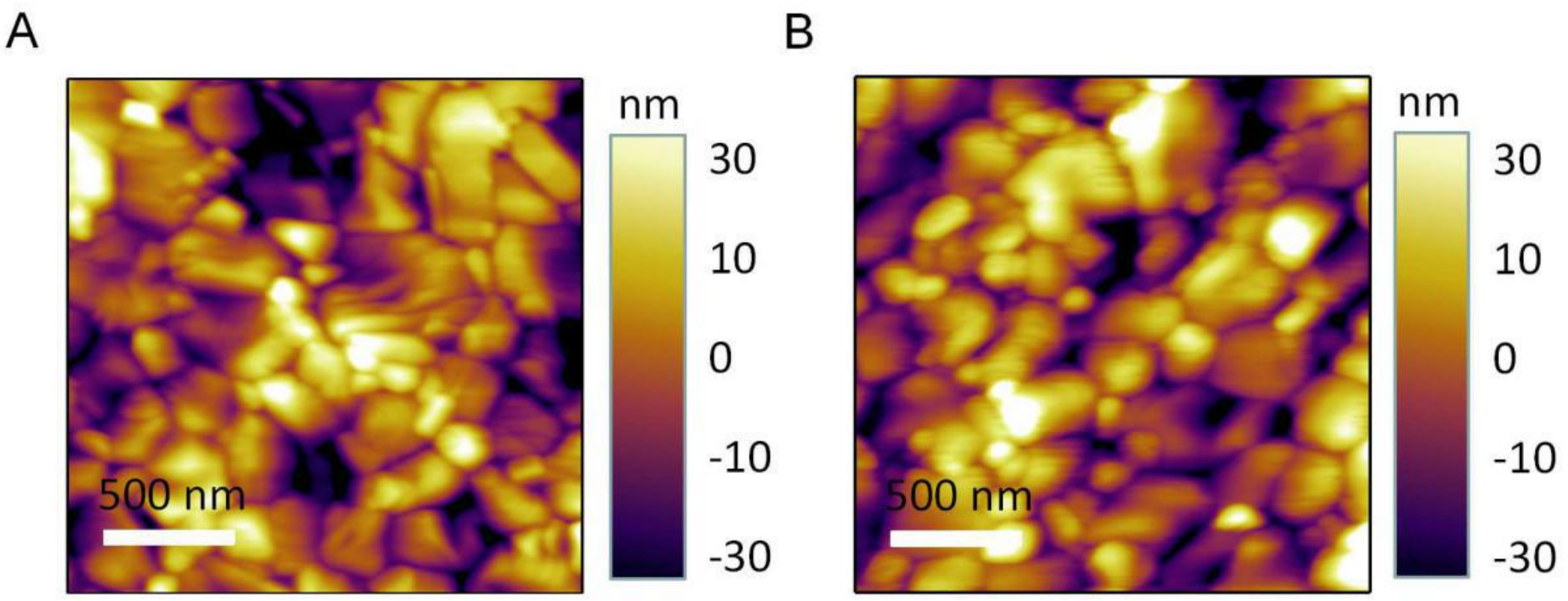

**Figure S14** The topographical of (A) PVK/PTAA/ITO and (B) KF-PVK-KF-PTAA-ITO samples from AFM measurements.

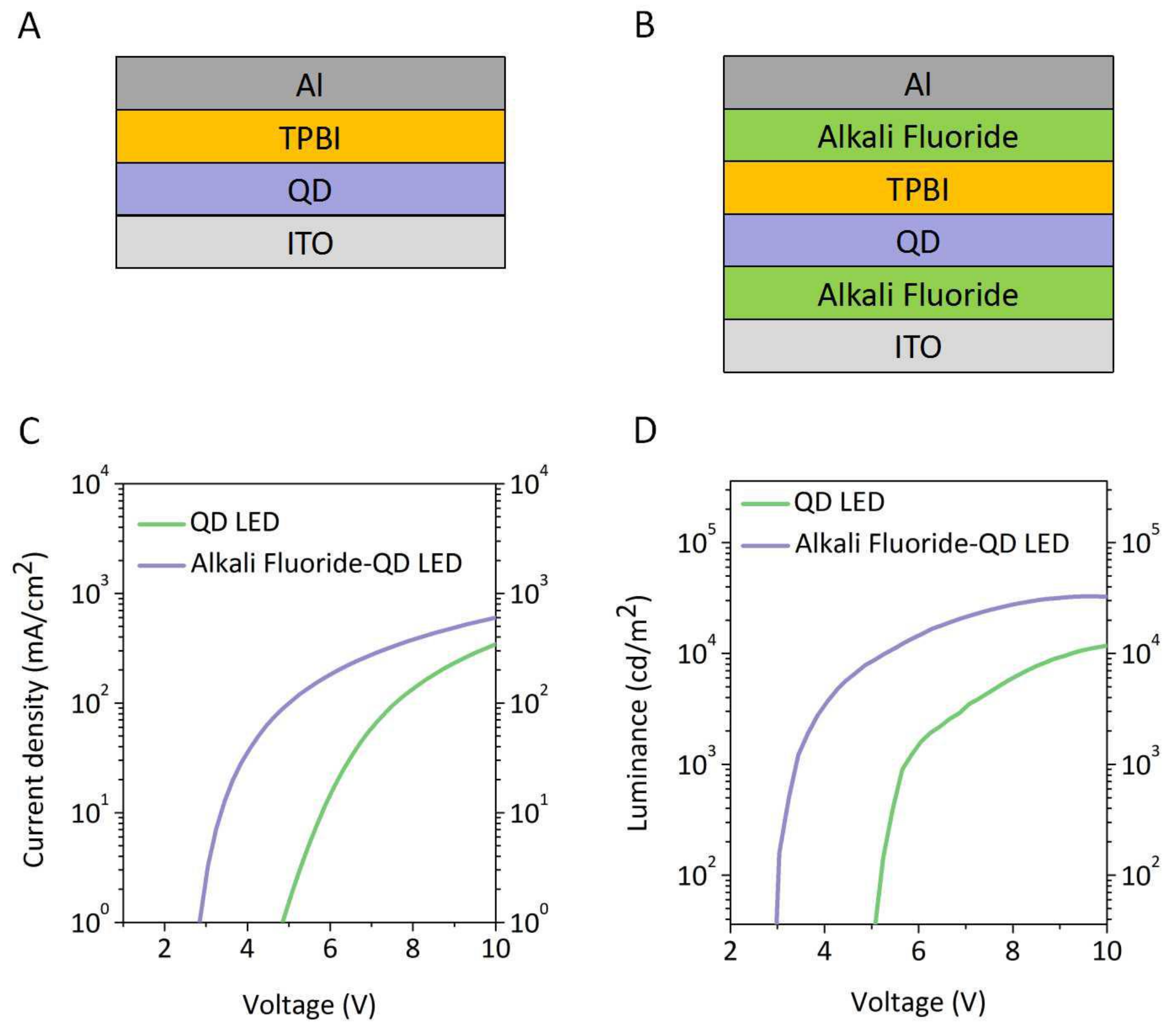

**Figure S15** Device structure of (A) the control perovskite quantum-dots light-emitting diode (QD LED) and (B) bilateral AF-modified perovskite QD LED (TPBI: 2,2',2 -(1,3,5-Benzinetriyl)-tris(1-phenyl-1-H-benzimidazole)). (C) Current-density and (D) luminance versus voltage curves for QD LEDs with (purple) and without (green) AF buffer layers between perovskite QDs and metal electrode.

**Table S5** Photovoltaic parameters for the control and the KF-perovskite-KF devices under indoor illumination.

| | Power density ($\mu W/cm^2$) | $J_{SC}$ ($mA/cm^2$) | $V_{OC}$ (V) | FF (%) | PCE (%) |
|---|---|---|---|---|---|
| Control | 292 | 0.126 | 0.81 | 71.2 | 24.9 |
| KF-perovskite-KF | 292 | 0.148 | 0.93 | 75.7 | 35.7 |

**Table S6** Photovoltaic performance for perovskite-based solar cells for indoor-light sources at 1000 lux.

| Structure | Power density ($\mu W/cm^2$ ) | Light Source/ Intensity (Lux) | $V_{OC}$ (V) | $J_{SC}$ ($mA/cm^2$) | FF (%) | PCE (%) | Refs |
|---|---|---|---|---|---|---|---|
| FTO/$NiO_X$/$CH_3NH_3PbI_3$/[BMIM]$BF_4$/Ag | 278.7 | 1000 | 0.87 | 0.150 | 75.2 | 35.2 | 8 |
| ITO/$NiO_X$/Perovskite/ PCBM/BCP/Ag | 160 | ~ | 0.88 | 0.059 | 69.6 | 21.4 | 9 |
| ITO/$NiO_X$/$I_3$-$I_2$Br-$I_{2-x}$$BrCl_x$/$PC_{601}$BM/BCP/Ag | 275.4 | 1000 | 1.0288 | 0.126 | 76.83 | 36.2 | 10 |
| ITO/PTAA/KF/Perovskite/KF/$C_{60}$/BCP/Ag | 292 | 1000 | 0.93 | 0.148 | 75.7 | 35.7 | This work |